\documentclass[twocolumn, tighten, twocolappendix]{aastex63} %anonymouslinenumbers linenumbers
\usepackage{pifont}
\usepackage{amsmath}

\newcommand{\cmark}{\ding{51}}%
\newcommand{\xmark}{\quad}%
\newcolumntype{P}[1]{>{\centering\arraybackslash}p{#1}}

\received{20th Sep 2022}
\revised{15th Mar 2023}
\accepted{???}
\shortauthors{Bonse et al.}
\begin{document}

\title{Comparing Apples with Apples: Robust Detection Limits for Exoplanet \\ High-Contrast Imaging in the Presence of non-Gaussian Noise}

\correspondingauthor{Markus J. Bonse}
\email{mbonse@phys.ethz.ch}

\author[0000-0003-2202-1745]{Markus J. Bonse}
\affiliation{ETH Zurich, Institute for Particle Physics \& Astrophysics, Wolfgang-Pauli-Str. 27, 8093 Zurich, Switzerland}
\affiliation{Max Planck Institute for Intelligent Systems, Max-Planck-Ring 4, 72076 T\"ubingen, Germany}

\author[0000-0003-2530-9330]{Emily O. Garvin}
\affiliation{ETH Zurich, Institute for Particle Physics \& Astrophysics, Wolfgang-Pauli-Str. 27, 8093 Zurich, Switzerland}

\author[0000-0001-9310-8579]{Timothy D. Gebhard}
\affiliation{ETH Zurich, Institute for Particle Physics \& Astrophysics, Wolfgang-Pauli-Str. 27, 8093 Zurich, Switzerland}
\affiliation{Max Planck Institute for Intelligent Systems, Max-Planck-Ring 4, 72076 T\"ubingen, Germany}
\affiliation{Max Planck ETH Center for Learning Systems, Max-Planck-Ring 4, 72076 T\"ubingen, Germany}

\author[0000-0002-5476-2663]{Felix A. Dannert}
\affiliation{ETH Zurich, Institute for Particle Physics \& Astrophysics, Wolfgang-Pauli-Str. 27, 8093 Zurich, Switzerland}
\affiliation{National Center of Competence in Research PlanetS (www.nccr-planets.ch)}

\author[0000-0002-3968-3780]{Faustine Cantalloube}
\affiliation{Aix Marseille University, CNRS, CNES, LAM, Marseille, France}

\author[0000-0001-7255-3251]{Gabriele Cugno}
\affiliation{ETH Zurich, Institute for Particle Physics \& Astrophysics, Wolfgang-Pauli-Str. 27, 8093 Zurich, Switzerland}
\affiliation{University of Michigan, Department of Astronomy, Ann Arbor, MI 48109, USA}

\author[0000-0002-4006-6237]{Olivier Absil}
\affiliation{Universit\'e de Li\`ege, STAR Institute, All\'ee du Six Ao\^{u}t 19c, B-4000 Li\`ege, Belgium}

\author[0000-0003-3768-5712]{Jean Hayoz}
\affiliation{ETH Zurich, Institute for Particle Physics \& Astrophysics, Wolfgang-Pauli-Str. 27, 8093 Zurich, Switzerland}
\affiliation{National Center of Competence in Research PlanetS (www.nccr-planets.ch)}

\author{Julien Milli}
\affiliation{Université Grenoble Alpes, CNRS, IPAG, 38000 Grenoble, France}

\author{Markus Kasper}
\affiliation{European Southern Observatory, Garching bei M\"unchen, Germany}

\author[0000-0003-3829-7412]{Sascha P. Quanz}
\affiliation{ETH Zurich, Institute for Particle Physics \& Astrophysics, Wolfgang-Pauli-Str. 27, 8093 Zurich, Switzerland}
\affiliation{National Center of Competence in Research PlanetS (www.nccr-planets.ch)}



\begin{abstract}
Over the past decade, hundreds of nights have been spent on the worlds largest telescopes to search for and directly detect new exoplanets using high-contrast imaging (HCI). Thereby, two scientific goals are of central interest: First, to study the characteristics of the underlying planet population and distinguish between different planet formation and evolution theories. Second, to find and characterize planets in our immediate Solar neighborhood. Both goals heavily rely on the metric used to quantify planet detections and non-detections. 
% How can we decide if a bright spot in our data is a planet or just noise? What are our detection limits? 
 %
 % Any error or inaccuracy made during the contrast estimation will directly affect all subsequent scientific reasoning.
 %
Current standards often rely on several explicit or implicit assumptions about the noise. For example, it is often assumed that the residual noise after data post-processing is Gaussian. While being an inseparable part of the metric, these assumptions are rarely verified. This is problematic as any violation of these assumptions can lead to systematic biases. This makes it hard, if not impossible, to compare results across datasets or instruments with different noise characteristics.
% This makes it hard, if not impossible, to compare results across datasets or instruments with different noise characteristics.
%
We revisit the fundamental question of how to quantify detection limits in HCI. We focus our analysis on the error budget resulting from violated assumptions. To this end, we propose a new metric based on bootstrapping that generalizes current standards to non-Gaussian noise. We apply our method to archival HCI data from the NACO-VLT instrument and derive detection limits for different types of noise. Our analysis shows that current standards tend to give detection limit that are about one magnitude too optimistic in the speckle-dominated regime. That is, HCI surveys may have excluded planets that can still exist.
\end{abstract}



% Extent structure from here on
\section{Introduction}
\label{sec:intro}
% We need a good metric to quantify our observations
In order to translate the information of exoplanet high-contrast imaging (HCI) observations into quantitative scientific results, an objective metric to quantify the brightness contrast between planet and host star is indispensable. In this context, two scientific questions are of particular interest:
% What a Standard is needed for - Two key research questions:
%
% Q1 - Is this real?
First, in case a new planet candidate is found, we want to know how confident we can be that the discovery is real. If several candidates are found during a survey, the calculated confidence can be used as a guideline for the management of follow-up time. 
%
% Q2 - What can we rule out?
Second, we want to know which planets we can safely rule out i.e. we are sure that, given some confidence level, we should have seen them in our data. This second question is the subject of studies about planet occurrence rates and completeness in large surveys such as SHINE \citep{desideraSPHEREInfraredSurvey2021,langloisSPHEREInfraredSurvey2021,viganSPHEREInfraredSurvey2021} or GPIES \citep{nielsenGeminiPlanetImager2019}.
Any metric used to answer these two questions must be able to deal with the large diversity of datasets available in HCI. Over the last years, HCI has gained increasing scientific importance as its discovery space complements other exoplanet detection and characterization techniques, such as radial velocity or transit observations. As a result, existing ground-based observatories feature, and continue to be upgraded with, dedicated HCI instruments \citep[e.g.,][]{kenworthyHighContrastImaging2018}. Furthermore, HCI exoplanet science is driving the development of future instruments for the upcoming 30-m class telescopes \citep[e.g,][]{quanzDirectDetectionExoplanets2015,bowensExoplanetsELTMETISEstimating2021,brandlMETISMidinfraredELT2021, kasperPCSRoadmapExoearth2021}. Also, the recently launched \emph{James Webb Space Telescope (JWST)} is equipped with a suite of HCI modes\footnote{\url{https://jwst-docs.stsci.edu/methods-and-roadmaps/jwst-high-contrast-imaging}} and a dedicated \emph{Early Release Science (ERS) program} was accepted for early execution \citep{hinkleyJWSTEarlyRelease2022}. In the long run, highly optimized space missions of large scale will be needed for the direct detection and characterization of a statistically relevant sample of temperate, terrestrial exoplanets \citep[e.g.,][]{gaudiHabitableExoplanetObservatory2018, quanzLargeInterferometerExoplanets2022, theluvoirteamLUVOIRMissionConcept2019}.

% Create the family picture
\begin{table*}[t]
	\caption{Standards for contrast quantification used in high-contrast imaging  \label{tab:family}}
	\begin{center}
	\begin{tabular}{p{4cm}P{0.5cm}P{0.5cm}P{1.5cm}P{2cm}P{1cm}P{2cm}P{3cm}}
	\hline
 	Method \& Reference & 
 	Q1 &
 	Q2&
 	Input type &
 	What is Noise? &
 	SSS &
 	Adresses Completeness & 
 	Accounts for non-Gaussian noise\\
 	\hline
 	\hline
 	 Estimation of noise PDF\tablenotemark{a} 
 		& \cmark & \cmark & RI & pixel & \xmark  & \xmark & \cmark\tablenotemark{y} \\
 		
 	 SNR \tablenotemark{b} 
 		& \cmark & \cmark & RI & both & \xmark  & \xmark &\xmark \\
 		
 	 t-Test (small sample)\tablenotemark{c}  
 		& \cmark & \cmark & RI & apertures & \cmark  & \xmark &\xmark \\
 		
 	Parametric P-Map \tablenotemark{d}   
 		& \xmark & \cmark & RI & apertures & \cmark & \cmark &\xmark \\
 		 		
 	Non-Parametric P-Map \tablenotemark{d}  
 		& \xmark & \cmark & RI & apertures & n.a.  & \cmark &\cmark\tablenotemark{z}  \\
 	
 	STIM \tablenotemark{e} 
 		& \cmark & \xmark & RI & pixel & \xmark   & \xmark & \cmark\tablenotemark{y} \\

 	ANDROMEDA\tablenotemark{f} 
 		& \cmark & \cmark & RS & pixel & \cmark  & \xmark & \xmark \\
 		
 	FMMF \tablenotemark{g} 
 		& \cmark & \cmark & RS  & pixel & \xmark  & \cmark & \cmark\tablenotemark{v} \\
 	
 	ROC curves \tablenotemark{h} 
 		& \xmark & \xmark & detection map \& RI & apertures or pixel & n.a.  & \xmark & \cmark\tablenotemark{z} \\
 		
 	Supervised ML \tablenotemark{i} 
 		& \cmark & \xmark & multiple RIs & pixel patches & n.a.  & \xmark  & \cmark\tablenotemark{u}  \\
 		
 	RSM \tablenotemark{j} 
 		& \cmark & \xmark & RS & apertures or pixel & n.a.  & \cmark  & \cmark\tablenotemark{w}  \\
 		
  	\hline
  	This paper
  		& \cmark & \cmark & RI  & spaced pixel & \cmark & \cmark\tablenotemark{x} & \cmark \\
  	\hline
\end{tabular}
	\end{center}
	\tablecomments{This table does not claim to be complete and is meant as an overview. We further note that the classification used here only considers a selection of the relevant aspects. A \cmark displays that the aspect is addressed in the cited paper, not including followup work. Abbreviations: %\newline
		\textbf{Q1}: Quantifies the uncertainty of a new detection,
		\textbf{Q2}: Provides detection limits,
		\textbf{SSS}: Accounts for small sample statistics,
		\textbf{RI}: Residual image,
		\textbf{RS}: Sequence of residuals along time,
		\textit{\textbf{u}}: Through learning typical noise pattern, \textit{\textbf{v}}: Estimates the distribution of the SNR over a whole survey, \textit{\textbf{w}}: Gaussian and Laplacian noise, \textit{\textbf{x}}: Future work, \textit{\textbf{y}}: Estimates the noise probability density function (PDF) based on pixel values, \textit{\textbf{z}}: No assumption about the noise.}
	\tablerefs{\quad 
		\mbox{\textit{a}: \cite{maroisConfidenceLevelSensitivity2008},}
		\mbox{\textit{b}: \cite{meshkatFURTHEREVIDENCEPLANETARY2013},}
	 	\mbox{\textit{c}: \cite{mawetFUNDAMENTALLIMITATIONSHIGH2014},}
	 	\mbox{\textit{d}: \cite{jensen-clemNewStandardAssessing2017},}
	 	\mbox{\textit{e}: \cite{pairetSTIMMapDetection2019},}
	 	\mbox{\textit{f}: \cite{cantalloubeDirectExoplanetDetection2015},}
	 	\mbox{\textit{g}: \cite{ruffioImprovingAssessingPlanet2017},}
	 	\mbox{\textit{h}: \cite{gonzalezLowrankSparseDecomposition2016},}
	 	\mbox{\textit{i}: \cite{gomezgonzalezSupervisedDetectionExoplanets2018},}
	 	\mbox{\textit{j}: \cite{dahlqvistRegimeswitchingModelDetection2020}}
	 	}
\end{table*}

Finding a metric that provides comparable estimates for contrast across current and future instruments is challenging because any change in instrumentation, observing conditions or data post-processing influences the characteristics of the data. Currently used standards often rely on several explicit or implicit assumptions about the noise. For example, it is often assumed that the noise in the post-processed residual image is independent identically distributed (i.i.d.) and Gaussian \cite[e.g.][]{mawetFUNDAMENTALLIMITATIONSHIGH2014,cantalloubeDirectExoplanetDetection2015}. While being an inseparable part of the metric, such assumptions are rarely verified. In fact, previous work has shown that the noise in ground-based HCI observations often deviates from Gaussian noise due to the presence of systematic speckle noise (compare \cite{perrinStructureHighStrehl2003, aimeUsefulnessLimitsCoronagraphy2004, soummerSpeckleNoiseDynamic2007} for the noise statistics in raw images and \cite{pairetSTIMMapDetection2019, dahlqvistRegimeswitchingModelDetection2020} for the noise statistics of post-processed residual images). 
%This special type of noise is caused by residual wavefront aberrations which remains uncorrected by the telescopes adaptive optics. 
If one computes detection limits under the assumption of Gaussian noise, but the actual noise is dominated by speckles, the results will be biased. The severity of this error depends on the extent to which the assumptions are violated. This makes it hard, or even impossible, to compare results across datasets and instruments with different noise characteristics.

% This work is about & Outline
In this paper, we revisit the fundamental question of \emph{"how to quantify detection limits"} in HCI. Compared to previous work \mbox{(Section \ref{sec:current_standards})}, we focus our analysis on the error budget caused by violated assumptions (\mbox{Section \ref{sec:critical}}). For this purpose we propose a new metric based on \emph{bootstrapping}, which generalizes the widely used standard by \cite{mawetFUNDAMENTALLIMITATIONSHIGH2014} \mbox{(Section \ref{sec:what_is_a_detection})} to non-Gaussian noise \mbox{(Section \ref{sec:beyond_gaussian_noise})}. This allows us to study the systematic error of our detection uncertainty as a function of the noise characteristics. We further highlight the important difference between the signal-to-noise ratio (SNR) and the detection uncertainty. In Section \ref{sec:quantify_detections_with_bs} we demonstrate how to quantify detections with our new bootstrapping metric. For this purpose, we present a new approach which is consistent with previous work but universally applicable irrespective of the algorithm used during data post-processing. We call this method the contrast grid. Finally, we analyze the error budget caused by non-Gaussian noise w.r.t. the detection limits in Section \ref{sec:results_and_discussion}.

The code of the metrics presented in this paper is public available as a \texttt{python} package called \texttt{applefy} on GitHub \url{https://github.com/markusbonse/applefy}. A detailed documentation page on how to use the package, including many tutorials, can be found at ReadtheDocs \url{https://applefy.readthedocs.io/}.
% ------------------------------------------------------------------------------------------------------------
\section{Current Standards for Contrast Quantification}
\label{sec:current_standards}
To this date, detection limits in HCI are not uniformly quantified, but are based on a wide range of standards (see  Table \ref{tab:family} for an overview). Many of these standards are used along with particular data reduction algorithms, which in turn are specialized for certain types of observing strategies. 
Most common observing strategies include angular differential imaging \citep[ADI][]{maroisAngularDifferentialImaging2006}, spectral differential imaging \citep[SDI][]{racineSpeckleNoiseDetection1999, sparksImagingSpectroscopyExtrasolar2002}, reference star differential imaging \citep[RDI][]{lafreniereHSTNICMOSDETECTION2009}, multi-reference star differential imaging \citep[mRDI][]{ruaneReferenceStarDifferential2019} and polarimetric differential imaging \citep[PDI][]{kuhnImagingPolarimetricObservations2001, quanzVERYLARGEESCOPE2011}. 

The data-reduction techniques can be categorized into three families \citep{cantalloubeExoplanetImagingData2021}: 
%
% PSF-subtraction
First, subtraction-based methods that try to model and subtract the stellar point spread function (PSF). Famous examples are PCA / KLIP \citep{amaraPYNPOINTImageProcessing2012, soummerDETECTIONCHARACTERIZATIONEXOPLANETS2012}, LOCI and its variations \citep{maroisExoplanetImagingLOCI2010, maroisGPIPSFSubtraction2014, wahhajImprovingSignaltonoiseDirect2015, thompsonImprovedContrastImages2021}, LLSG \citep{gonzalezLowrankSparseDecomposition2016} and the recently proposed HSR \citep{gebhardHalfsiblingRegressionMeets2022}. The output of these techniques is a \emph{residual image} whose values are related to flux. The quantification of the detection limits is carried out in a separate step. 
%
% Forward modeling
Second, forward-modeling techniques aim at tracking potential planetary signals during the observing sequence. Their result is not a residual, but a \emph{detection map} showing the model's belief about the presence of a planet. Contrast and detection uncertainy are defined as the match of expected signal and the data. Forward-modeling for HCI was introduced in \cite{cantalloubeDirectExoplanetDetection2015} with the ANDROMEDA algorithm and extended in various ways; see, for example FMMF \citep{ruffioImprovingAssessingPlanet2017}, PACO \citep{flasseurExoplanetDetectionAngular2018} and TRAP \citep{samlandTRAPTemporalSystematics2021}. 
Third, supervised machine learning (ML) methods \citep{gomezgonzalezSupervisedDetectionExoplanets2018} try to learn the specific signatures of signal and noise in the data. Their output is again a \emph{detection map}. 

Because numerous methods are used to post-process the data, the metrics differ fundamentally in their definition of what a detection is. This starts with the \emph{research question} they address: Some metrics focus only on the detection problem, that is whether a potential candidate is real (Q1). Other metrics try to constrain detection limits (Q2). Yet other methods specialize solely on comparing post-processing algorithms \citep[e.g. ROC-curves][]{gonzalezLowrankSparseDecomposition2016}. 
%
% A metric in this context is a measure for the reliability of planet detection.
%
Even more fundamentally, there is no uniform standard to define \emph{what is signal and what is noise}. For example, some methods calculate the strength of the noise based on areas around the signal \citep{maroisConfidenceLevelSensitivity2008, mesaPerformanceVLTPlanet2015, ottenONSKYPERFORMANCEANALYSIS2017, golombPlanetEvidencePlanetNoise2021} while others consider noise with the same distance from the star \citep{cantalloubeDirectExoplanetDetection2015, mawetFUNDAMENTALLIMITATIONSHIGH2014, jensen-clemNewStandardAssessing2017}. Other variants use the opposite angle rotated residual \citep{pairetSTIMMapDetection2019, wahhajGEMININICIPLANETFINDING2013} or estimates along time \citep{dahlqvistRegimeswitchingModelDetection2020}. The noise statistics is sometimes calculated directly on pixel values \citep{maroisConfidenceLevelSensitivity2008, mesaPerformanceVLTPlanet2015, cantalloubeDirectExoplanetDetection2015, ruffioImprovingAssessingPlanet2017}, while others average multiple pixel inside apertures \citep{mawetFUNDAMENTALLIMITATIONSHIGH2014, jensen-clemNewStandardAssessing2017}.
Moreover, different types of \emph{statistics} are used. While some authors use classical definitions of signal-to-noise ratios \citep{rameauDISCOVERYPROBABLE452013, meshkatFURTHEREVIDENCEPLANETARY2013, mesaPerformanceVLTPlanet2015, uyamaSEEDSHighContrastImaging2017}, others address the quantification using frequentist hypothesis testing \citep{mawetFUNDAMENTALLIMITATIONSHIGH2014, jensen-clemNewStandardAssessing2017} or Bayesian methods \citep{ruffioBayesianFrameworkExoplanet2018, golombPlanetEvidencePlanetNoise2021}.

% We have to place this figure here in oder to have it on top of the page covering the detection topic
\begin{figure*}[t!]
	\epsscale{1.2}
	\plotone{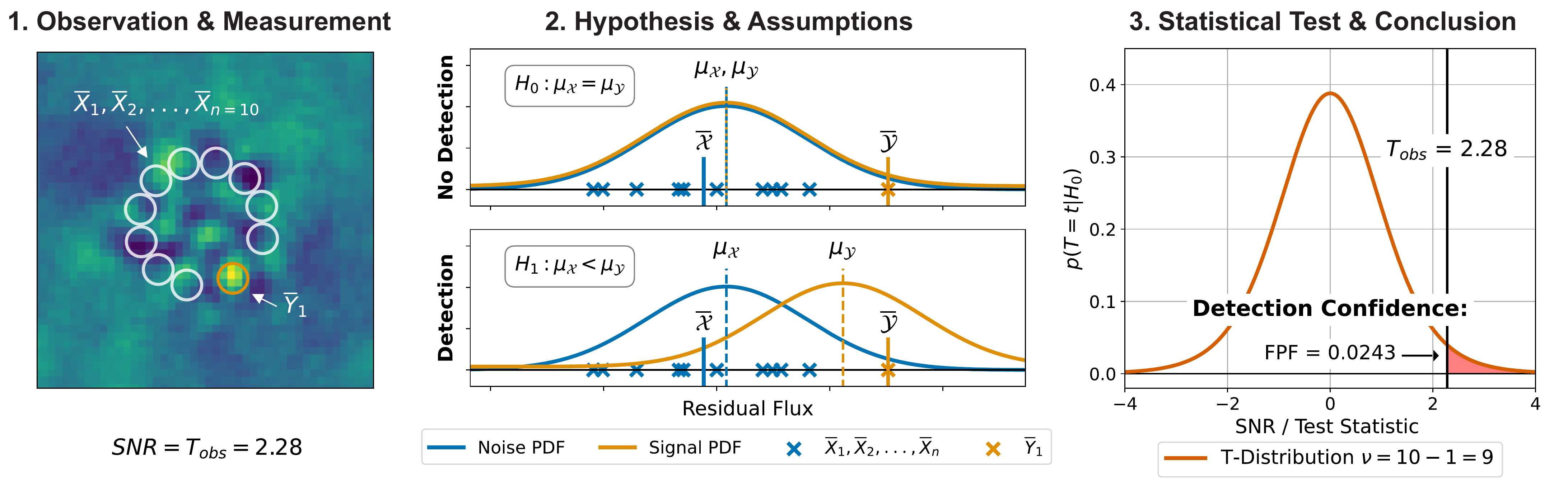}
	\caption{The three main steps of the hypothesis testing framework introduced by \cite{mawetFUNDAMENTALLIMITATIONSHIGH2014} which we generalize in this paper. \textit{Step 1} shows a typical example of a residual image obtained after data post-processing with PCA. A previously inserted artificial companion appears as a spot slightly brighter than the remaining speckle noise. More details on the dataset are given in \mbox{Section \ref{sec:critical}}. In order to determine if the planet is real we extract values for noise $X$ and signal $Y$ by averaging pixel values inside apertures. In \textit{Step 2} we formulate two competing hypotheses $H_0$ (top) and $H_1$ (bottom) for detection and non-detection, respectively. The orange and blue crosses correspond to the signal and noise values obtained from the residual image on the left. Is the signal bright enough to show that $\mathcal{X}$ and $\mathcal{Y}$ differ significantly in their means? In order to answer this question we calculate the test statistic of the two sample t-test $T_{\text{obs}}=2.28$ (Equation \ref{eq:SNR}). \textit{Step 3}: The detection uncertainty or false-positive-fraction (FPF) is given by the shaded red area under the t-distribution with $\nu = n-1 =9$ degrees of freedom.}
	\label{fig:hypothesis_testing}
\end{figure*}

Due to these differences, results calculated with different standards are often not comparable. 
%The same applies to all scientific analyses founded upon the calculated detection limits, e.g. planetary occurrence rates. 
But which of the presented metrics gives the "right" scientific answers? On the one hand, an ideal metric should be \emph{universally applicable} irrespective of the observing strategy or data reduction used. On the other hand, it should be \emph{robust} under different data characteristics and provide reliable estimates for the achieved contrast. 
% The metric should not facilitate more detections or higher contrast.
%
% Limits of current standards
Current standards are limited with respect to both criteria: First, they are often specialized to be used with specific post-processing algorithms. For example, FMMF is used after post-processing with PCA.
Second, they are reliant on fixed assumptions about the noise making them non-robust under varying conditions.

In the following \mbox{Section \ref{sec:what_is_a_detection}} we revisit the metric presented in \cite{mawetFUNDAMENTALLIMITATIONSHIGH2014} and assess its limitations. Afterwards, we propose modifications and extensions to this approach in order to improve the robustness and universal applicability of the metric.

\section{What is a Detection?} 
\label{sec:what_is_a_detection}

% General Background
The data post-processing routine combines the sequence of individual observations taken over the course of one night into a single image. In this residual image potential planet candidates appear as bright spots. A priori, we do not know if a bright spot is actually a planet, and the decision of whether it is bright enough to be counted as a detection is always a balancing of the risk that it is just part of the noise.
In order to quantify this risk, we use hypothesis testing. The classical approach based on the t-test \citep[for a general background on hypothesis testing and the t-test see Chapter 9 of][]{larsenIntroductionMathematicalStatistics2012} was introduced by \cite{mawetFUNDAMENTALLIMITATIONSHIGH2014} and is illustrated in Figure \ref{fig:hypothesis_testing}. The test procedure can be split into three main steps:

% Step 1:
In the first step, we extract values for the noise and the potential planet. It is important that the noise is taken from positions that are representative for the noise we expect at the position of the signal. A common approach is to use noise with the same separation from the star. Further, noise and signal must be extracted in the same way, ideally independent of the resolution of the detector.
This can be done by averaging pixel values inside apertures of $1 \lambda / D$ diameter. The result are two samples: the noise sample $\mathcal{X} = \{\overline{X}_1, ..., \overline{X}_n \}$ and the planet sample $\mathcal{Y} = \{\overline{Y}_1 \}$. Note, that the planet sample only contains a single observation.

% Step 2:
% two sample test in general
% Introduce the general problem: We have observations: 1. a planet sample 2. a noise sample; We want to make a statement about the underlaying distributions' expected values.
In the next step, two competing hypotheses are formulated: first, that our observation can be explained without the presence of a planet. This is the \emph{null hypothesis} $H_0$. Second, the hypothesis that our observation is indeed attributable to the existence of a planet. This is the \emph{alternative hypothesis} $H_1$. 
%
% Test for differences in means -> signal present vs signal absent
% The concrete Version in Mawet et al
In order to decide which of the two hypotheses to favor we need to make assumptions and explicitly formulate $H_0$ and $H_1$. The t-test assumes that our observations are drawn independent and identically distributed (i.i.d) from normal distributions $\{\overline{X}_1, ..., \overline{X}_n \} \sim \mathcal{N}(\mu_\mathcal{X}, \sigma_\mathcal{X})$, $\{\overline{Y}_1\} \sim \mathcal{N}(\mu_\mathcal{Y}, \sigma_\mathcal{Y})$ where the parameters $\mu_\mathcal{X}, \mu_\mathcal{Y}, \sigma_\mathcal{X}, \sigma_\mathcal{Y}$ are unknown. In addition, it is assumed that the distributions of the signal and the noise have the same variance\footnote{
% why the assumption about similar variance might be valid
Note, that the value of $\overline{Y}_{1}$ is not just the pure planet signal but also contains speckle noise: $\overline{Y}_{1} = S_{\text{planet}} + \overline{X}_{n+1}$ where $S_{\text{planet}}$ is the contribution of the planet signal and $\overline{X}_{n+1}$ is the speckle noise at the position of the planet. We can describe $S_{\text{planet}}$ as Poisson (photon shot noise) and keep the Gaussian assumption for $\overline{X}_{n+1}$. The resulting probability density function (PDF) for $\overline{Y}_{1}$ is then the convolution of the Poisson PDF with the Gaussian PDF \citep[Theorem 5.2.9 in][]{casellaStatisticalInference2002}. Thus, by assuming equal variance, we assume that the effect of photon shot noise is negligible compared to the speckle noise.} $\sigma_\mathcal{X} = \sigma_\mathcal{Y}=\sigma$.
In case no planet is present we would expect that $\mathcal{X}$ and $\mathcal{Y}$ originate from the same distribution. Thus, we can formulate $H_0: \mu_\mathcal{X} = \mu_\mathcal{Y}$. In case a planet is present we expect $\mathcal{Y}$ to be brighter than $\mathcal{X}$. This gives us $H_1: \mu_\mathcal{X} < \mu_\mathcal{Y}$. 
%
%Note, any difference in means greater than zero $\delta_\mu = \mu_Y - \mu_X >0$ is counted as a detection. 
%
% What influences our certainty about the differences in means?
A closer look at step 2 in Figure \ref{fig:hypothesis_testing} shows that $H_0$ becomes unlikely as the distance between the noise observations (blue crosses) and planet observation (orange cross) increases and the variance of the noise sample decreases.
%
% Step 3:
We can quantify this effect using the \textit{test statistic} of the two sample t-test \citep{mawetFUNDAMENTALLIMITATIONSHIGH2014}

\begin{equation}
\label{eq:SNR}
	T = \frac{\hat{\mu}_\mathcal{Y} - \hat{\mu}_\mathcal{X}}{\hat{\sigma}_{\mathcal{X}}\sqrt{1 + 1/n}} \,.
\end{equation}
The equation is simplified for the special case in HCI where $\mathcal{Y}$ contains only a single value. The mean estimates of the signal and noise sample are denoted as $\hat{\mu}_\mathcal{Y}=\overline{\mathcal{Y}} = \overline{Y}_1$ and $\hat{\mu}_\mathcal{X}=\overline{\mathcal{X}}$, while $\hat{\sigma}_\mathcal{X}$ is the Bessel corrected standard deviation of $\mathcal{X}$.
%The pooled sample standard deviation $\hat{\sigma}_{X, Y}$ can be calculated by:
%\begin{equation}
%	\hat{\sigma}^2_{X, Y} =  \frac{(n-1)\hat{\sigma}_X^2 + (m-1)\hat{\sigma}_Y^2}{n+m-2}
%\end{equation}
%
Under $H_0$ and if $\hat{\mu}_\mathcal{X}$ and $\hat{\mu}_\mathcal{Y}$ follow a normal distribution, $T$ follows a student t-distribution with $\nu=n-1$ degrees of freedom. The t-distribution does not depend on the unknown parameters $\mu_\mathcal{X}, \mu_\mathcal{Y}, \sigma$. This allows us to compute the \textit{detection uncertainty} by

\begin{equation}
	\label{eq:FPF_ttest}
	\text{FPF} = \int_{T_{\text{obs}}}^{\infty} p(T = t | H_0) \,d x \,,
\end{equation}
where $T_{\text{obs}}$ denotes the value of $T$ that we compute for our observation and $p(T = t | H_0)$ is the probability density function of the t-distribution.
The false-positive-fraction (FPF) gives the risk that we reject $H_0$ in favor of $H_1$ although no planet is present. In panel 3 of \mbox{Figure \ref{fig:hypothesis_testing}} we obtain $T_{\text{obs}} = 2.28$ which corresponds to an $\text{FPF}=0.0243$. That is, in $2.43\%$ of the cases in which no planet is present we get a $T_{\text{obs}} \geq 2.28$. If the calculated FPF is below a previously defined detection threshold we treat our observation as a detection. Since small values of the FPF quickly become difficult to read, we use the following notation in this paper:
\begin{equation}
	x \, \sigma_{\mathcal{N}} := 1 - \Phi(x) = \text{FPF}
\end{equation}
where $\Phi$ is the cumulative density function of the standard normal distribution. That is, we express the FPF values in terms of the quantiles of the standard normal distribution $x$. For example we write $5\sigma_{\mathcal{N}}$ for $\text{FPF}=2.87 \times 10^{-7}$ and $3\sigma_{\mathcal{N}}$ for $\text{FPF}=1.25 \times 10^{-3}$. Note, $\sigma_{\mathcal{N}}$ should not to be confused with $\hat{\sigma}_{\mathcal{X}}$.

% Difference between noise and statistics distribution
It is crucial to distinguish between the assumed distribution of the noise (Gaussian) and the distribution of the test statistic (t-distribution). The t-distribution does not describe the nature of the noise but the effect of the sample size. 
For small $n$ the values of $\hat{\mu}_\mathcal{X}$,  $\hat{\mu}_\mathcal{Y}$ and $\hat{\sigma}_{\mathcal{X}}$ are less accurate w.r.t. the true but unknown parameters $\mu_\mathcal{X}$,  $\mu_\mathcal{Y}$ and $\sigma$. This uncertainty is compensated by the heavier tails of the t-distribution for small $\nu$. 
%
% relationship to classical SNR
For $n \rightarrow \infty$ Equation \ref{eq:SNR} converges to the classical definition of the signal-to-noise ratio (SNR) \citep[compare e.g.][]{meshkatFURTHEREVIDENCEPLANETARY2013} and the t-distribution converges to the normal distribution. 
In this limit we obtain the classical false alarm probabilities of $\text{FPF}=2.87 \times 10^{-7}$ for $T=5$ and $\text{FPF}=1.25 \times 10^{-3}$ for $T=3$. 
%
% SNR vs FPF: Why one should report confidence levels not SNR
Note that the factor $1 / \sqrt{1 + 1/n}$ in Equation \ref{eq:SNR} is not a correction for the classical SNR in case of small sample sizes. It is a normalization that ensures that $p(T = t | H_0)$ follows a standard t-distribution. The small sample size at inner radii affects both, the value of $T_{\text{obs}}$ as well as the shape of the t-distribution. Any value of $T_{\text{obs}}$ or SNR is meaningless if we do not consider the sample size and underlying  assumptions of the test. For example at $2 \lambda /D$ we need $ T_{\text{obs}}\geq 11.2$ for a $5\sigma_{\mathcal{N}}$ detection while at $10 \lambda /D$ a value of $ T_{\text{obs}}\geq5.6$ is sufficient. Therefore, any detection threshold should be specified as a FPF and not in terms of $T_{\text{obs}}$ nor SNR. 

\section{Critical Assessment of Assumptions}
\label{sec:critical}
% General motivation for the critical assessment 
In the following, we use the t-test described in the previous section as an example to critically assess the validity of the assumptions used in HCI. We note, that some of the metrics shown in \mbox{Table \ref{tab:family}} are based on assumptions similar as those of the t-test. For example the assumed noise distribution in forward modelling techniques defines the  maximum likelihood expression which is the basis for the calculation of the detection map. Thus, although our analysis is based on the t-test, some aspects discussed in this paper might be of general interest. 
Apart from the assumptions used in HCI, some implementation details can have a non-negligible effect on the results. One example is the placement of the apertures in the residual image. We cover this aspect in \mbox{Appendix \ref{sec:rotation}} and focus our discussion on the two main assumptions of the t-test: the independence and the Gaussianity of the residual noise. 

% The description of the dataset here
Throughout the rest of this paper we compute our results based on a \mbox{$L'$-dataset} of \mbox{$\beta$ Pictoris} taken with the AGPM coronagraph \citep{delacroixLaboratoryDemonstrationMidinfrared2013} and NACO \citep{roussetNAOSFirstAO2003} at the VLT. The dataset is the same as the one used in \cite{mawetFUNDAMENTALLIMITATIONSHIGH2014}. The planet \mbox{$\beta$ Pictoris b} was removed by insertion of a negative fake planet. More details on the data can be found in \cite{absilSearchingCompanionsAU2013}. Our data pre-processing routine uses the \texttt{python} library PynPoint given in \cite{stolkerPynPointModularPipeline2019}. The final stack of pre-processed frames consists of 29681 images.

% Aspect 2: Independence ------------------------------------------------------------------------
\subsection{Independence}
\label{sec:independence}
A central assumption of the t-test is that $\mathcal{X}$ and $\mathcal{Y}$ are drawn independently. This means that all noise values have to be uncorrelated with respect to all other noise values and the signal. 
%
% Different noise for different instruments
Depending on the observing strategy, instrument, wavelength and data reduction, the noise characteristics of HCI observations can be fundamentally different. For example, observations at 3-5$\mu m$ are often dominated by the thermal background noise whereas observations at shorter wavelength 1-2.5 $\mu m$ are more affected by speckles \citep{hunzikerPCAbasedApproachSubtracting2018}. The noise type also affects the spatial correlations in the data. While some noise sources such as photon noise are independent on a pixel-to-pixel level, the planet signal and speckle noise will follow the shape of the telescopes PSF. 
If the noise is dominated by speckles and the pixel size is smaller than the width of the PSF, neighboring pixel are not independent. Some methods listed in \mbox{Table \ref{tab:family}} compute their noise statistics directly on pixel values while assuming independence. In the presence of speckle noise these methods will not provide accurate FPF estimates.

% Figure about independence
\begin{figure}[t]
	\epsscale{1.}
	\plotone{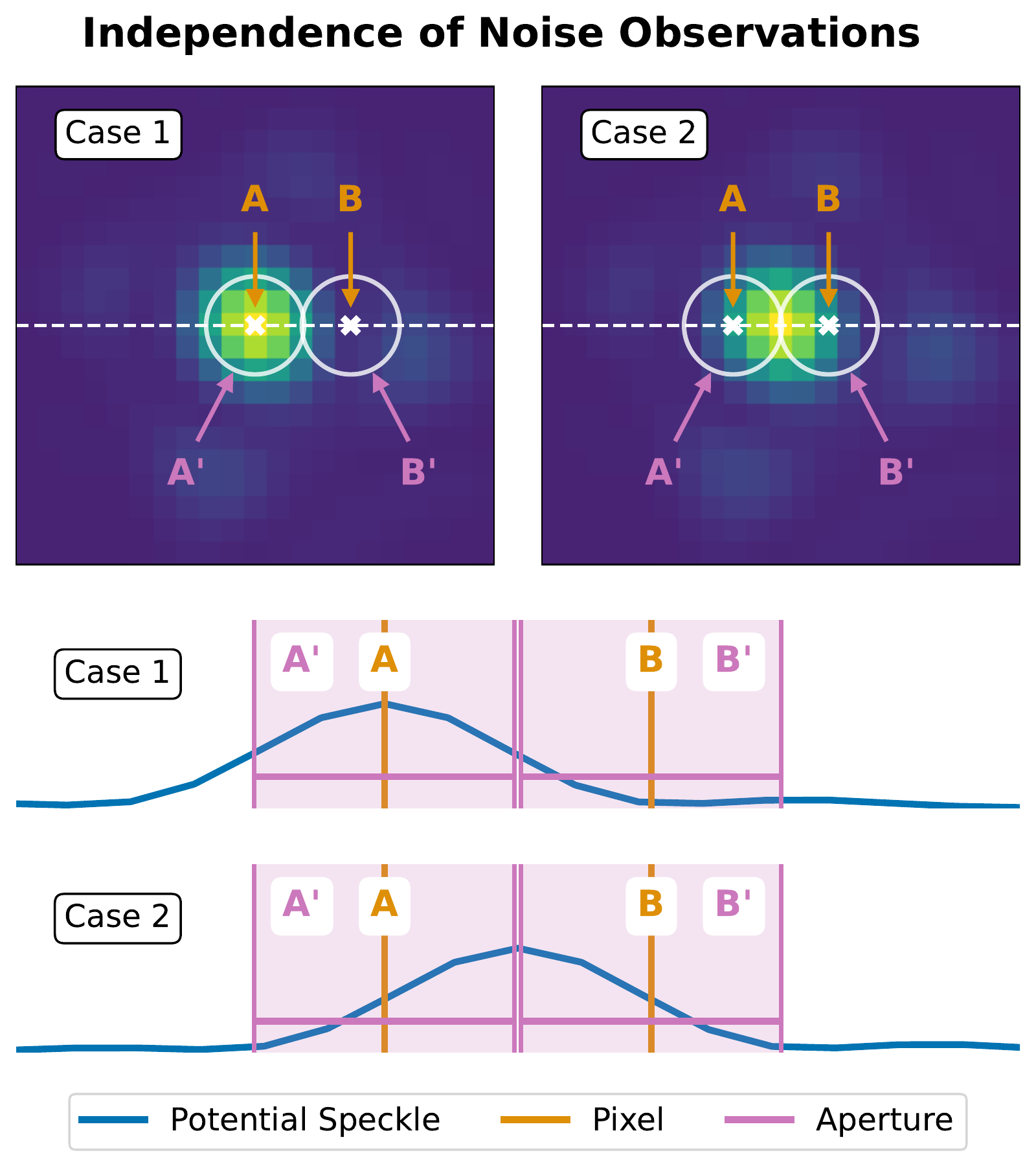}
	\caption{Violations of the independence assumption in the presence of speckle noise. The top two images illustrate the two cases described in the text. The pixel $A$ and $B$ are separated by 1 FWHM. We use the unsaturated PSF of the NACO dataset to display the speckle. The plots below give one dimensional profiles of the two cases, cut along the white dashed lines. The shaded areas highlight the information contributing to the aperture averages $A'$ and $B'$.}
	\label{fig:independence}
\end{figure}
Compared to an analysis directly on the pixels, the use of apertures suggest that $\mathcal{X}$ and $\mathcal{Y}$ are independent. A careful investigation, however, reveals that this impression is incorrect. Let us consider two pixel $A$ and $B$ together with a speckle on the detector (compare \mbox{Figure \ref{fig:independence}}). If the two pixel are distant by less than \mbox{1 FWHM} they will be partly correlated as both are influenced by the same speckle. This is the case if the speckle is located at $A$ (case 1) or between $A$ and $B$ (case 2). Therefore, if we calculate aperture averages $A'$ and $B'$ around $A$ and $B$ respectively, they will always be based on non-independent pixel values.
If we use apertures we implicitly filter the residual image with a box filter which has the shape of the aperture. Consequently, the length of the spatial correlations increases and with it the risk for a violation of the independence assumption.

We therefore propose to use the pixel values at the positions $A$ and $B$ directly instead of apertures for datasets in the speckle dominated regime. The separation between $A$ and $B$ should be chosen according to the expected spatial correlation length in the data. 
As bad seeing conditions or poor adaptive optics performance can influence the shape of the PSF, we space $A$ and $B$ by one FWHM and do not use the theoretical size of $\lambda /D$. The FWHM can be calculated by fitting a 2D Gaussian or Moffat to the unsaturated PSF \citep{stolkerPynPointModularPipeline2019}. The FWHM for the $\beta$ Pictoris dataset is 4.2 pixel and slightly larger than the theoretical size in $\lambda /D$ \cite[see also][]{jensen-clemNewStandardAssessing2017}. We note, that spacing $A$ and $B$ by one FWHM does not guarantee that their values are completely independent (compare case 2). But their values will be less dependent compared to the values of $A'$ and $B'$.
In addition to speckles other effects might influence the spatial dependencies in the data. A common example for this is the wind-driven halo discussed in \citep{cantalloubeWinddrivenHaloHighcontrast2020}. In such cases high-pass spatial filtering can be used to recover the spatial independence of the noise. The use of low-pass filters such as the Gaussian blur \citep{absilSearchingCompanionsAU2013}, however, exacerbates the independence problem. 
We further note, that in reality the spatial correlations can deviate from the shape of the PSF e.g. due to data post-processing. An empirical analysis on this topic is given in \mbox{Appendix \ref{sec:Correlation}}. 

In the background-dominated regime the use of apertures might be preferable over the use of spaced pixels. Since, the photon noise occurs on a pixel-by-pixel basis it is not problematic with respect to the independence. But, the signal estimate based on the brightest pixel is prone to statistical fluctuations and biases such as hot pixels. The choice whether apertures or spaced pixel are used should be taken on a case by case basis which is why our \texttt{python} package \texttt{applefy} provides an implementation of both.

Future work should further investigate more sophisticated alternatives to the use of spaced pixel including explicit models of the pixel-to-pixel dependencies \citep[compare e.g.][]{golombPlanetEvidencePlanetNoise2021}. 

% Aspect 3: The noise is not Gaussian. --------------------------------------------------------------
\subsection{The assumption of Gaussian noise}
\label{sec:non_gaussian_noise}
\begin{figure}[t]
\epsscale{1.15}
\plotone{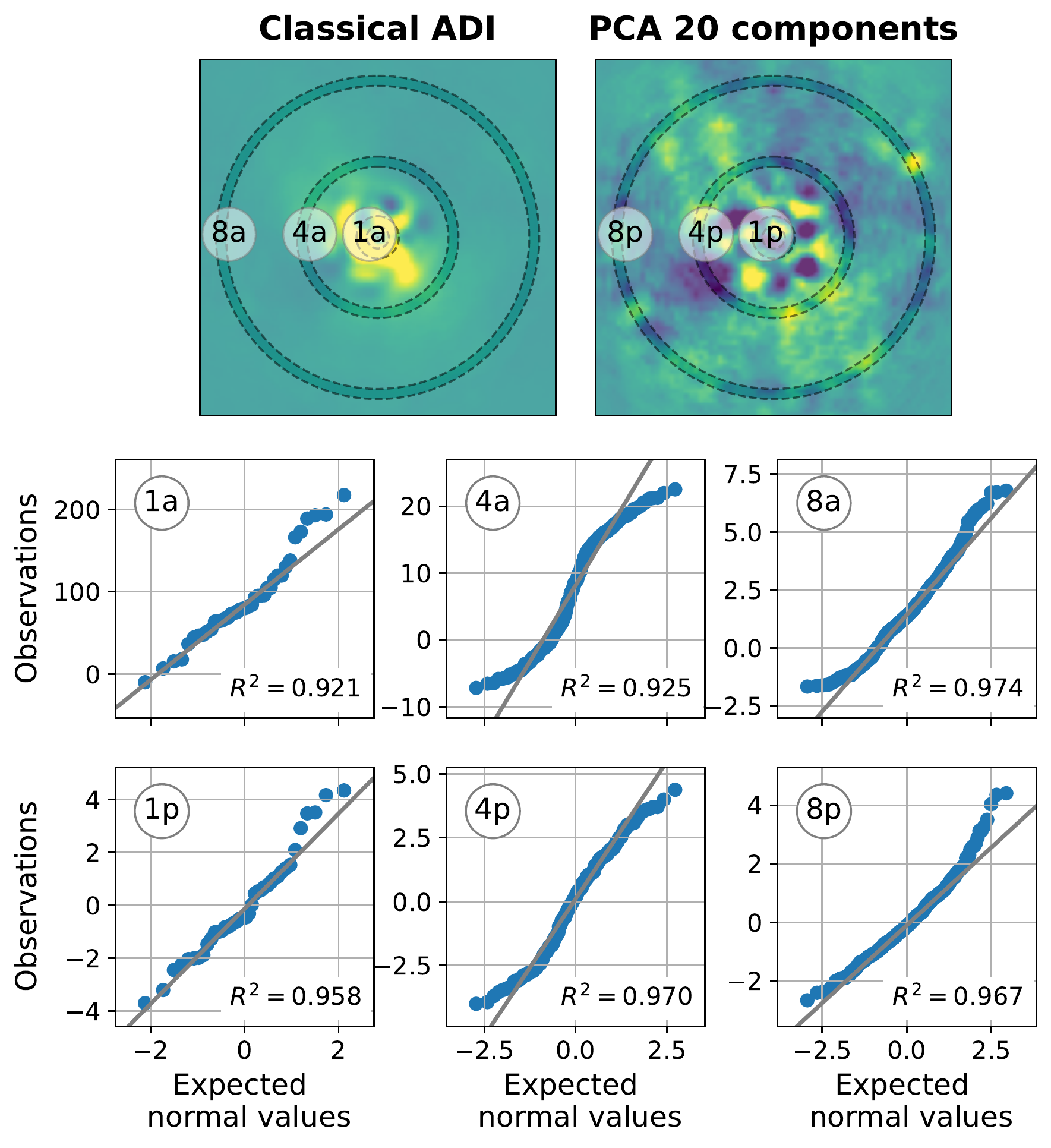}
\caption{Deviations from the normal distribution in HCI residuals. The top two images show residual images for the {$\beta$ Pictoris} dataset: Left with classical median ADI \citep{maroisAngularDifferentialImaging2006}; right with full-frame PCA and 20 components. Below Q-Q plots are given to study the noise statistic of the pixel values inside the shaded areas at \mbox{1 FWHM}, \mbox{4 FWHM} and \mbox{8 FWHM}. The Q-Q plot compares the observed pixel values with normal distributed noise. A perfect match of observations and normal distribution would result in points exclusively on the grey diagonal line. We note, that the pixel values extracted from the shaded areas are not independent. The shown Q-Q plots only provide indicative evidence for the type of residual noise. But, they are not a proof for or against Gaussian distributed noise. A discussion on the topic can be found in \mbox{Appendix \ref{sec:testing_gaussian}}.
%The grey line shows the best fit between observed and theoretical noise using a Theil–Sen linear regression \citep{theil1950rank,sen1968estimates}. $R^2$ is the coefficient of determination.
}
\label{fig:residual_statistics}
\end{figure}

% The Gaussian assumption
The t-test assumes that the noise and the signal average $\overline{\mathcal{X}}, \overline{\mathcal{Y}}$ follow a normal distribution. This assumption is usually justified by the use of observing strategies such as ADI \citep{maroisAngularDifferentialImaging2006}, which average many individual observations. In this way, a temporal sequence of sufficiently i.i.d observations will yield a residual image with normal distributed noise by virtue of the central limit theorem (CLT) \citep{maroisConfidenceLevelSensitivity2008, mawetFUNDAMENTALLIMITATIONSHIGH2014}. In addition, the use of data post-processing techniques, such as a PSF subtraction with PCA, have demonstrated to considerably improve the normality of residuals \citep{amaraPYNPOINTImageProcessing2012, soummerDETECTIONCHARACTERIZATIONEXOPLANETS2012, cantalloubeDirectExoplanetDetection2015}. 
Despite the frame averaging and data post-processing, residual noise of real observations can still deviate from Gaussian. In \mbox{Figure \ref{fig:residual_statistics}} we use Q-Q plots\footnote{Q-Q plots are a statistical tool to compare the quantiles of two distributions with each other. For a detailed explanation, see \cite{pairetSTIMMapDetection2019} and references therein.}
 to compare the noise of the \mbox{$\beta$ Pictoris} dataset with normally distributed noise. 
At 1 FWHM and 8 FWHM \mbox{(labels 1 \& 8)} we notice that the noise for large values is above the diagonal line of the Q-Q plot. This implies that bright pixel are more frequent in the data than we expect from Gaussian noise. Similar noise statistics were previously observed by \cite{maroisConfidenceLevelSensitivity2008,cantalloubeDirectExoplanetDetection2015,pairetSTIMMapDetection2019}. At 4 FWHM \mbox{(labels 4a and 4p)} we observe the opposite behavior: large values are less frequent in our data compared to the normal distribution.
A measure for these deviations is the coefficient of determination $R^2$ that is the Pearson correlation between the paired sample quantiles. See \cite{pairetSTIMMapDetection2019} for the definition and detailed explanation. The closer $R^2$ is to 1, the better the observed noise can be explained by Gaussian noise. As suggested by the results shown in \mbox{Figure \ref{fig:residual_statistics}} the use of PCA partially mitigates the problem of non-Gaussian noise. Nevertheless, the noise is still not perfectly normal.

% Why is the deviation a problem?
In the presence of noise which has a high probability of large values to occur, the probability that we observe a large value of $T$ (\mbox{Equation \ref{eq:SNR}}) that is caused by the noise increases. Consequently, $p(T = t | H_0)$  will no longer follow a t-distribution and the interpretability of the test statistic $T$ w.r.t. the FPF is lost. That is, we can still calculate values for $T$, but we no longer know which detection uncertainty they are associated with. Depending on the type of noise, different values of $T_{\text{obs}}$ might be required to reject $H_0$. This is especially problematic as noise characteristics can change from dataset to dataset. If we ignore potential violations of the Gaussian assumption, we under- or over-estimate the detection uncertainty. 
%
% What are the limits for n vs 1 -> infinity?
% Why does the problem still matter at large separations / CLT does not help
For many applications outside HCI, the t-test is robust to slightly non-Gaussian data given a large sample size. As the sample size increases the average values of $\overline{\mathcal{X}},\overline{\mathcal{Y}}$ will be Gaussian thanks to the CLT. In HCI, we cannot take advantage of this effect as $\mathcal{Y}$ always contains a single observation. For small separations, where speckle noise is most important, the sample size of $\mathcal{X}$ is also limited.
How strong is the effect of non-Gaussian noise on the detection uncertainty? Is a comparison between datasets or instruments with different noise characteristics still possible? 

\section{Beyond Gaussian Noise}
\label{sec:beyond_gaussian_noise}
\begin{figure*}[t!]
	\epsscale{1.05}
	\plotone{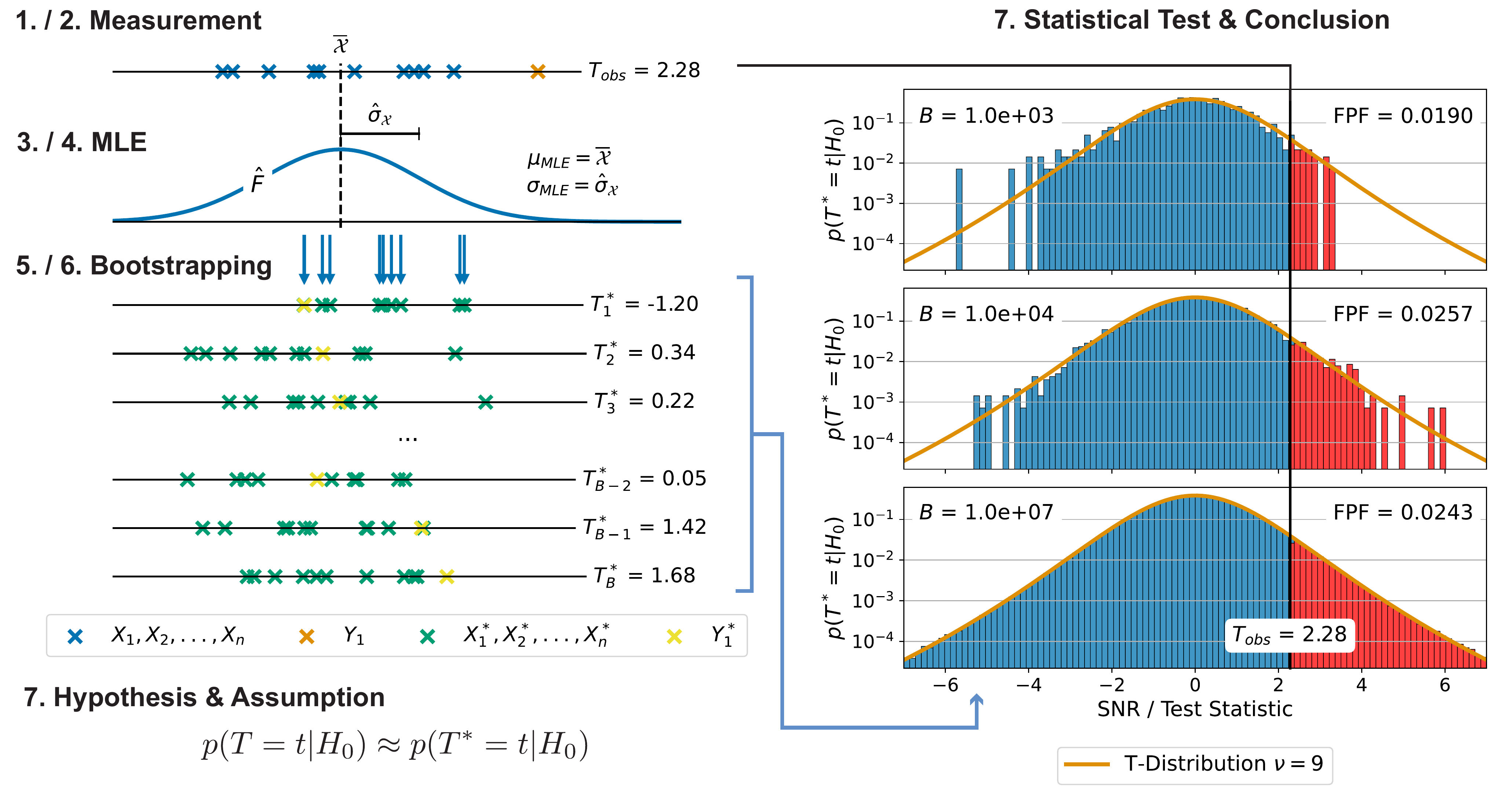}
	\caption{The parametric bootstrap test introduced in this paper. The left side of the figure illustrates the key steps of the test, exemplary for Gaussian noise. The values used are identical to those in Figure \mbox{\ref{fig:hypothesis_testing}}. The numbers of the steps match with those used in the text. On the right, three histograms are given, which show the empirical distribution of $p(T^* = t | H_0)$ for different numbers of bootstrap samples $B$. The actual test does not require to compute histograms and the FPF is computed from the bootstrap samples \mbox{$T_{(1)}^*\leq  ... \leq T_{(B)}^*$} directly (see \mbox{Equations \ref{eq:fpf_bs_simple} and \ref{eq:FPF_para_bs}}). The results are compared with the t-distribution with $\nu = n -1 = 9$ degrees of freedom. Compared to the t-test, bootstrapping is much more versatile, since any distribution, not necessary Gaussian, can be assumed in step 3.}
	\label{fig:para_bs_workflow}
\end{figure*}

% How to fix the assumption about Gaussian noise:
In general, we can distinguish three concepts to deal with non-Gaussian noise:
%
% Find a better assumption
First, we can assume a different noise distribution which better describes the speckle behavior. The RSM presented in \cite{dahlqvistRegimeswitchingModelDetection2020} explores this idea, though along the time domain and not in the residual image. 
%
% Estimate the noise from the data
Second, we can try to estimate the distribution of the noise directly from the data. This idea was previously studied by \cite{maroisConfidenceLevelSensitivity2008} and \cite{pairetSTIMMapDetection2019} who tried to estimate the noise probability density function (PDF) based on pixel values in the residuals. As mentioned before, this can be problematic since the pixels in the residuals are not independent. Further, extensive extrapolation is required in order to describe how the noise behaves in the tails of the PDF. The concept is therefore better suited for large surveys as e.g. done by \cite{ruffioImprovingAssessingPlanet2017}.
%
% No assumption about the noise
Finally, we can try to estimate detection limits by not making any assumptions about the noise. An example for this are ROC-curves \citep{gonzalezLowrankSparseDecomposition2016, jensen-clemNewStandardAssessing2017} which are often used to benchmark post-processing algorithms.

In the following, we propose a new metric based on bootstrapping which follows the first concept and allows us to compute the detection uncertainty for any type of noise. The other two concepts are subject of future work.
In order to retain the interpretability of the test statistic $T$, we need to find its distribution $p(T = t | H_0)$ . In principle this could be done empirically. That is, we repeat our observation a lot of times on identical stars that do not host a planet and compute values for $T$ for every observation. The probability $p(T = t | H_0)$  is then given by the frequency that a value $T$ is observed. 
While repeating the observation that many times is not feasible in practice, it gives the key idea of the bootstrap: we aim at finding a distribution $\hat{F}$ that is close to the true but unknown distribution of the noise $F$ and use it to \emph{repeat the experiment}.
Two main concepts can be distinguished: first, the non-parametric bootstrap which makes no further assumptions about the noise and tries to approximate $F$ directly from the data. Second, the parametric bootstrap which assumes that $F$ follows a parametric distribution with unknown parameters.
In practice, if only a few noise observations are available, the estimated noise model $\hat{F}$ of the non-parametric bootstrap might be inaccurate w.r.t. the true $F$. This can lead to less accurate estimates of the test statistic distribution $p(T = t | H_0)$ .
In this paper we study how to compute detection uncertainties using the parametric bootstrap, that is, under the assumption that the noise distribution is known. The data-driven non-parametric bootstrap will be explored in future work.

\subsection{The Parametric Bootstrap}
\label{sec:para_bs}
% The steps of the parametric bootstrap used in this paper
We adopted the bootstrap test discussed in example 3.4.1 of \cite{zoubirBootstrapTechniquesSignal2004} and chapter 16.2 of \cite{efronIntroductionBootstrap1993} to the special case of HCI where only a single value for $\mathcal{Y}$ is available. The test is similar to the two sample t-test explained in \mbox{Section \ref{sec:what_is_a_detection}} except that we can assume any noise distribution. The following procedure describes the main steps of the test, which are the same irrespective of the assumed noise distribution. For demonstration purposes, we give examples of how the test is carried out in the case of Gaussian noise (see \mbox{Figure \ref{fig:para_bs_workflow}}):
%
%If the assumed noise distribution satisfies some additional properties, the bootstrap procedure can be simplified significantly. We will discuss this topic in \mbox{Section \ref{sec:pivot}}. 
%The main steps of the parametric bootstrap test are:
%
\begin{enumerate}
	\item Extract the noise sample $\mathcal{X} = \{X_1, ..., X_n\}$ and the signal sample $\mathcal{Y} = \{Y_1\}$ from the residual image\footnote{We write $\mathcal{X} = \{X_1, ..., X_n\}$ and $\mathcal{Y} = \{Y_1\}$ instead of $\mathcal{X} = \{\overline{X}_1, ..., \overline{X}_n\}$ and $\mathcal{Y} = \{\overline{Y}_1\}$ to emphasize that we use spaced pixel instead of apertures (see discussion in \mbox{Section \ref{sec:independence}}).}. As for the t-test we assume that the noise is independent and identically distributed.
	\item Compute $T_{obs}$ by using \mbox{Equation \ref{eq:SNR}}.
	\item Assume a parametric distribution for the noise and list its unknown parameters. In case of Gaussian noise this would be $F=\mathcal{N}(\mu_\mathcal{X}, \sigma)$.
	\item Compute the maximum likelihood estimate of the unknown parameters. For Gaussian noise we compute the mean $\hat{\mu}_\mathcal{X} = \overline{\mathcal{X}}$ and the standard deviation $\hat{\sigma} = \text{std}(\mathcal{X})$ and set $\hat{F}=\mathcal{N}(\hat{\mu}_\mathcal{X}, \hat{\sigma})$. 
	\item \emph{Repeat the experiment} $B$ times under $H_0$ using numerical simulation based on $\hat{F}$. For every repetition we resample the noise \mbox{$\mathcal{X}^* = \{X_1^*, ..., X_n^*\} \sim \hat{F}$} and the signal \mbox{$\mathcal{Y}^* = \{Y_1^*\} \sim \hat{F}$} from the same distribution. It is important to draw the same number of noise observations for $\mathcal{X}^*$ as we have in $\mathcal{X}$. The results are $B$ so called bootstrap samples 
	\begin{eqnarray*}
		&(\{X_1^*, ..., X_n^*\}, \{Y_1^*\})_1, ..., (\{X_1^*, ..., X_n^*\}, \{Y_1^*\})_B \\ &= (\mathcal{X}^*_1, \mathcal{Y}^*_1), ... , (\mathcal{X}^*_B, \mathcal{Y}^*_B)
	\end{eqnarray*}
	\item For every bootstrap sample $(\mathcal{X}^*_i, \mathcal{Y}^*_i)$ compute the test statistic $T_i^*$ by using \mbox{Equation \ref{eq:SNR}}. Rank the results with increasing order \mbox{$T_{(1)}^*\leq  ... \leq T_{(B)}^*$}.
	\item Under the assumption that $\hat{F}$ is close to $F$ the distribution of $p(T^*=t | H_0)$ will converge to $p(T = t | H_0)$  if $B$ becomes large. The convergence depends on the maximum likelihood estimates calculated in step 4. If this estimate is inaccurate the bootstrap results will be biased. However, if the assumed noise distribution satisfies some additional properties explained in \mbox{Section \ref{sec:pivot}}, we are able to obtain the exact distribution of \mbox{$p(T = t | H_0)$}  irrespective of how close $\hat{F}$ is to $F$. 
	
	The false positive fraction is given by the fraction of values in $\{T_1^*, ..., T_B^*\}$ which are larger than $T_{\text{obs}}$:
	\begin{equation}
		\label{eq:fpf_bs_simple}
		\text{FPF} = \#\{T_i^* > T_\text{obs}\} / B \, .
	\end{equation}
	More accurate results can be obtained by linear interpolation. For this purpose we use the sorted bootstrap results \mbox{$T_{(1)}^*\leq  ... \leq T_{(B)}^*$} from the previous step and search for the two values of $T^*$ which are adjacent to $T_{\text{obs}}$: $T^*_{(a)} \leq T_{\text{obs}} \leq T^*_{(b)}$, where $a$ and $b$ are the corresponding indices. The detection uncertainty can be computed with  
	\begin{eqnarray}
	\label{eq:FPF_para_bs}
	1 -\text{FPF} = &\frac{a}{B-1} \cdot \frac{T_{\text{obs}} - T^*_{(a)}}{T^*_{(b)} - T^*_{(a)}} \\ \nonumber
				    + &\frac{b}{B-1} \cdot \frac{T^*_{(b)} - T_{\text{obs}}}{T^*_{(b)} - T^*_{(a)}} \, .
	\end{eqnarray}
\end{enumerate}
As shown in \mbox{Figure \ref{fig:para_bs_workflow}} the calculated FPF of the bootstrap test converges to $\text{FPF} = 0.0243$. This is exactly the same value that we obtained with the t-test in \mbox{Section \ref{sec:what_is_a_detection}}. This means that the parametric bootstrap is equivalent to the t-test if we assume Gaussian noise.

\subsection{The Parametric Bootstrap for Laplacian Noise}
\label{sec:para_bs_laplace}
\begin{figure}[t]
	\epsscale{1.12}
	\plotone{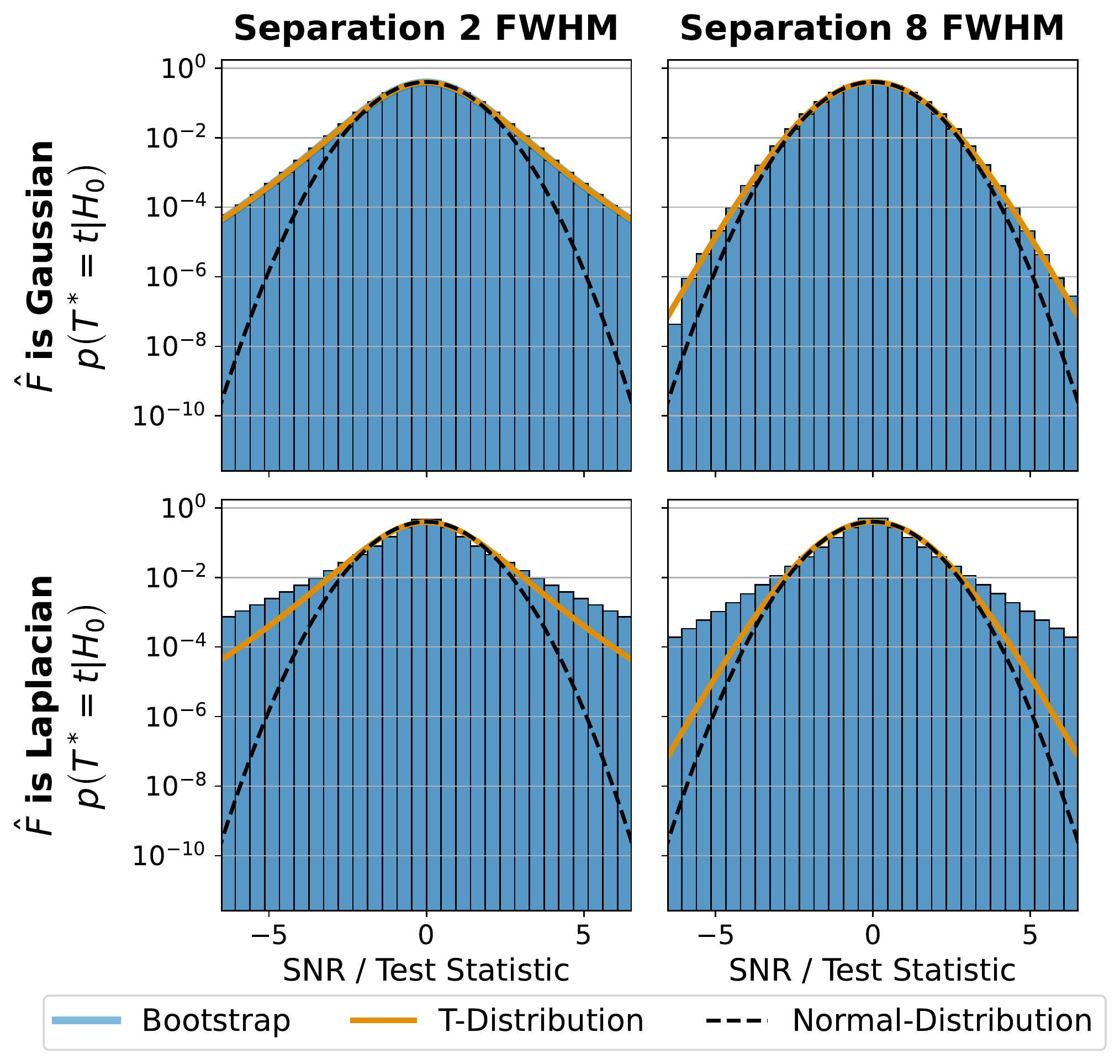}
	\caption{The convergence distributions of the parametric bootstrap, shown as histograms, in comparison with the normal and t-distribution. The top two plots show $p(T = t | H_0)$  under the assumption that $F$ is Gaussian, once at a separation of $2 \, \text{FWHM}$ i.e. $n=11$ and once at $8 \, \text{FWHM}$ i.e. $n=49$. The plots show the same result as the right side of \mbox{Figure \ref{fig:para_bs_workflow}} but for different $n$. The bottom two plots show $p(T = t | H_0)$  under the assumption of Laplacian noise. We use $B=10^{8}$ bootstrap iterations.}
	\label{fig:bs_convergence}
\end{figure}

In contrast to the t-test, the parametric bootstrap allows us to compute the FPF for any type of noise. This can be done by swapping out the Gaussian assumption in step 3.
% Example for the Laplace
In coronagraphic images with small static residuals, the modified Rician distribution of the speckle noise \citep{soummerSpeckleNoiseDynamic2007,aimeUsefulnessLimitsCoronagraphy2004} is reduced to a one-sided exponential \citep{fitzgeraldSpeckleStatisticsAdaptively2006}. This applies for example in the case of coronagraphic imaging with small non-common path aberrations or after some basic data post-processing. If we subtract two such images from each other, as done in the ANDROMEDA algorithm \citep{cantalloubeDirectExoplanetDetection2015}, we expect the noise to follow a two sided-exponential distribution i.e. a Laplacian. Similarly, if a PCA-model tries to subtract bright speckles, it sometimes erroneously induces negative speckles leading to a similar type of noise. Even after averaging several frames along the temporal dimension, the residual noise of some HCI datasets is often better described\footnote{Often it is not possible to find sufficient evidence that the residual noise is indeed Laplacian or Gaussian. More details on the topic are given in Appendix \ref{sec:testing_gaussian}.} by Laplacian and not by Gaussian noise \citep[see][for a detailed analysis on this topic]{pairetSTIMMapDetection2019}. 
In order to extend the bootstrap procedure to Laplacian noise we first assume $F = \mathcal{L}(\mu_{\mathcal{X}}, b)$ with PDF:

\begin{equation}
	\label{eq:laplace_pdf}
	f(x| \mu_{\mathcal{X}}, b) = \frac{1}{2b}\exp \left(-\frac{|x - \mu_{\mathcal{X}}|}{b}\right) \, .
\end{equation}
The maximum likelihood parameters \mbox{(step 4)} for $\mu_{\mathcal{X}}$ and $b$ are given by \citep{nortonDoubleExponentialDistribution1984}:
\begin{eqnarray}
	\hat{\mu}_{\mathcal{X}} = &\text{median}(\mathcal{X}) \\
	\hat{b} = &\frac{1}{n} \sum_{i=1}^{n} |X_i - \hat{\mu}_{\mathcal{X}}|
\end{eqnarray}
This gives us  $\hat{F} = \mathcal{L}(\hat{\mu}_{\mathcal{X}}, \hat{b})$ which we use to repeat the experiment by resampling (compare left side of \mbox{Figure \ref{fig:para_bs_workflow}}).
The remaining steps of the test are identical to those in \mbox{Section \ref{sec:para_bs}}\footnote{
The test statistic in \mbox{Equation \ref{eq:SNR}} is not necessarily optimal for all types of noise. It might be possible to derive a new test statistic for specific situations which still gives correct FPF estimates and at the same time offers higher power. This could be done by utilizing the Neyman-Pearson lemma \citep[Theorem 3.1.1 in][]{zoubirBootstrapTechniquesSignal2004}. We leave this idea open for future work and focus our analysis on the calculation of the error budget using the test statistic in \mbox{Equation \ref{eq:SNR}}.}.
By changing $F$ from a Gaussian to a Laplacian, we are able to compute the FPF under the assumption of Laplacian noise and the distribution of the test statistic $p(T = t | H_0)$  (compare right side of \mbox{Figure \ref{fig:para_bs_workflow}}) will no longer be a t-distribution. A comparison of $p(T = t | H_0)$  under the assumption of Gaussian and Laplacian noise is given in \mbox{Figure \ref{fig:bs_convergence}}.
If the assumed noise distribution is Gaussian, the distribution of $p(T = t | H_0)$  resulting from bootstrapping agrees with the t-distribution. This applies to separations close to the star as well as further out. If the assumed noise distribution is Laplacian, the bootstrap converges to a distribution with even heavier tails than the t-distribution. These heavy tails allow us to correct for the high occurrence of large noise values in Laplacian noise. We note, the distributions shown in \mbox{Figure \ref{fig:bs_convergence}} are not the distributions of the noise (here Gaussian and Laplacian), but the distribution of the test statistic $p(T = t | H_0)$ . 

In step 5 of the bootstrap algorithm we sample the same number of noise observations $n$ as in the original \mbox{sample $\mathcal{X}$}. This allows us to simultaneously consider the effects of the small sample statistics and non-Gaussian noise. The procedure can be extended to any parameterized noise \mbox{distribution $F$} for which the maximum likelihood estimates in step 4 are known.

\subsection{The Importance of Pivoting}
\label{sec:pivot}
In general, if we make no further restrictions on the type of noise distribution assumed, the accuracy of the bootstrap test depends on the maximum likelihood estimates calculated in step 4 in \mbox{Section \ref{sec:para_bs}}. Any deviation of the estimated $\hat{F}$ from the true $F$ will change \mbox{$p(T^*=t | H_0)$} and with it bias the FPF. However, for some specific types of noise distributions, we can overcome this limitation by taking advantage of so-called pivotal quantities \citep[see definition 9.2.6 in][]{casellaStatisticalInference2002}: "a random variable $Q(\mathcal{X}, \theta)$ is a pivotal quantity if the distribution of $Q(\mathcal{X}, \theta)$ is independent of all parameters. That is, if $\mathcal{X} \sim F(x|\theta)$, then $Q(\mathcal{X}, \theta)$ has the same distribution for all values $\theta$."

% Explain what this means in case of the t-test and Gaussian noise
In case of the t-test the noise and planet sample follow a normal distribution $\{X_1, ..., X_n \} \sim \mathcal{N}(\mu_\mathcal{X}, \sigma)$, \mbox{$\{Y_1\} \sim \mathcal{N}(\mu_\mathcal{Y}, \sigma)$} with unknown parameters $\theta=\{\mu_{\mathcal{X}}, \mu_{\mathcal{Y}}$, $\sigma\}$. Under the null hypothesis, the test statistic $T$ \mbox{ in Equation \ref{eq:SNR}} is a pivotal quantity. That is, the distribution of $p(T = t | H_0)$  is always the same, irrespective of the unknown parameters $\mu_{\mathcal{X}}, \mu_{\mathcal{Y}}, \sigma$. This means that no matter which maximum likelihood estimates $\hat{\mu}_{\mathcal{X}}$ and $\hat{\sigma}$ we calculate, the bootstrap procedure will always converge to the same distribution. For the t-test this distribution is the t-distribution. The fact that $T$ is a pivot is crucial for the t-test: it allows us to compute the exact FPF based on the t-distribution without knowing $\mu_{\mathcal{X}}, \mu_{\mathcal{Y}}$ and $\sigma$. 

In \mbox{Appendix \ref{sec:pivot_proof}} we prove that $T$ is a pivot not only under the assumption of Gaussian noise but for all distributions from a location-scale family. A location-scale family is characterized by two unknown parameters: a scale parameter $q$ and a shift parameter $w$. Let $Z \sim \mathcal{F}$ be a random variable following a standard distribution with no unknown parameters. By shifting and scaling $Z$ we obtain the location-scale family with random variables $X = Z q + w$. In case of the normal distribution $Z$ is the standard normal distribution $Z \sim \mathcal{N}(0, 1)$ with $q = \sigma, w=\mu$. Also the Laplacian distribution is a location-scale family were $Z \sim \mathcal{L}(0, 1)$ with PDF

\begin{equation}
	f(x) = \frac{1}{2}\exp \left(-|x|\right) 
\end{equation}
and $q = b, w=\mu$ (compare \mbox{Equation \ref{eq:laplace_pdf}}). 
The result that $T$ under $H_0$ is a pivotal quantity for location-scale family distributions has direct implications for the bootstrap procedure explained in \mbox{Section \ref{sec:para_bs}}:
The computation of the maximum likelihood parameters in step 4 is no longer needed. The resampling in steps 5 and 6 is independent of $\hat{\mu}_{\mathcal{X}}$ and $\hat{\sigma}$ (or $\hat{b}$) and always converges to the same $p(T^*=t | H_0)$. No matter how close $\hat{\mu}_{\mathcal{X}}$ and $\hat{\sigma}$ (or $\hat{b}$) are to the true values of $\mu_{\mathcal{X}}$ and $\sigma$ (or $b$), we always obtain the exact FPF.
The estimation of the FPF becomes a simple lookup. We can calculate $p(T^*=t | H_0)$ once and reuse it for future calculations, which drastically reduces the computation time.

The distribution of $p(T^*=t | H_0)$ depends on the sample size $n$ which is why we have to compute separate lookups for different separations from the star. In order to achieve accurate results for low FPF, the number of bootstrap samples $B$ has to be large. For example, if we aim for a FPF of $2.87 \cdot 10^{-7}$ ($5\sigma_{\mathcal{N}}$), we expect to observe one false-positive every \mbox{$B=1/2.87 \cdot 10^{-7} = 3.48 \cdot 10^{6}$} iterations. To constrain $p(T^*=t | H_0)$ with sufficient accuracy, we need about $B=10^8$ bootstrap samples.
In our \texttt{python} package \texttt{applefy} we provide pre-computed lookups within $1 - 20 \, \lambda /D$ for $B=10^8$ and interfaces which can be used to extend the code to other noise distributions.

\section{How to quantify detections with parametric bootstrapping?}
\label{sec:quantify_detections_with_bs}
Current standards for contrast quantification in HCI (compare \mbox{Section \ref{sec:current_standards}}) address two main research questions: First, the detection problem, i.e. whether a potential candidate is real (Q1), and second, the quantification of the detection limits (Q2). The following two subsections give a detailed recipe of how parametric bootstrapping can be used to address these two questions. An implementation is given in our \texttt{python} package \texttt{applefy}.

\subsection{How to compute the Uncertainty of a Detection?}
\label{sec:how_to_detection}
In \mbox{Section \ref{sec:critical}} we identified two limitations of the t-test when used in HCI: 1. The violation of the independence assumption if we use apertures  2. The non-Gaussian residual noise. To overcome these limitations, we propose the following procedure to quantify whether a signal in the data is a real planet candidate:
\begin{enumerate}
	\item Make an assumption about the type of noise present in the residual. For a conservative choice and in the case of speckle noise we recommend to choose Laplacian noise over Gaussian noise. If the noise is from a location scale family, use parametric bootstrapping to pre-compute the distribution of the test statistic under the null hypothesis $p(T^*=t | H_0)$ (see \mbox{Section \ref{sec:pivot}}). In case of Gaussian noise use the t-distribution. For surveys the use of the non-parametric bootstrap briefly mentioned in \mbox{Section \ref{sec:beyond_gaussian_noise}} represents an alternative to a fixed assumption about the noise. 
	
	\item The planet sample $\mathcal{Y}$ and noise sample $\mathcal{X}$ need to be extracted in the same way. Since we use spaced pixel for the noise, we have to use one pixel as the planet signal to get commensurable quantities. We search for the position of the planet by fitting a 2D Gaussian. The flux integrated over the circular area of one pixel around the best fit position gives us the planet signal $\mathcal{Y}$.	
	
	\item Extract noise values $\mathcal{X}$ with the same separation from the star. To ensure that the values in $\mathcal{X}$ are approximately independent, we do not use apertures but integrated over the circular area of one pixel spaced by one FWHM. In the background limited regime the use of apertures might be preferable for this and the previous step.

	\item Use \mbox{Equation \ref{eq:SNR}} to compute the test statistic $T_{\text{obs}}$\footnote{The assumption made in step 1 only influences the distribution of the test statistic $p(T = t | H_0)$ . The test statistic itself is the same.}.
	
	\item Translate $T_{\text{obs}}$ into the detection uncertainty (FPF) by using \mbox{Equation \ref{eq:FPF_ttest}} or \mbox{Equation \ref{eq:FPF_para_bs}}.
	
	\item Repeat steps 3. to 5. for different noise positions. This can be done by rotation of the initial noise positions (compare discussion in \mbox{Appendix \ref{sec:rotation}}  and \mbox{Figure \ref{fig:rotation}}).
	
	\item Report the median FPF over all noise positions. The mean absolute deviation from the median can be used as a measure of the uncertainties introduced by the placement of the noise values.
\end{enumerate}

\subsection{How to compute Detection Limits?}
\label{sec:detection_limits}
If no planet candidate is found, we can compute detection limits to constrain which planets existence we can confidently rule out. To this end, artificial planets are usually inserted in order to determine the minimum planet brightness still detectable \citep{maroisExoplanetImagingLOCI2010, morzinskiMAGELLANADAPTIVEOPTICS2015}. Hereby, the calculation of the detection limit has to be consistent with the calculation of the FPF in the previous section. That is, if we insert an artificial planet above the detection limit we expect it to be counted as a detection. Vice versa, if we insert an artificial planet below the detection limit it should not be detectable. 

In reality detection limits are never a hard limit and only give the point at which approximately 50\% of the planets are detected. There is no guarantee that all planets above the contrast curve will be detected and no planet below the curve will not be detected. At the position of the planet we always observe a combination of planet signal and speckle noise. 
If we are lucky and a faint planet falls on top of a speckle, we might be able to detect it although it is below our calculated limits. Conversely, if a negative speckle caused by the data post-processing is added to the planet signal, it may fall below the detection threshold. This effect can be quantified through calculations of the true-positive rate (TPR) also called the power of the test. A detailed discussion about the topic is presented in \cite{jensen-clemNewStandardAssessing2017}. In this paper we focus only on the FPF and leave power calculations for the parametric bootstrap open for future work. 

In the following we introduce a new approach to compute detection limits which is independent of the data post-processing. We call this approach the \emph{contrast grid}.
The calculation starts with the insertion of artificial planets at different separations e.g. \mbox{$s = 1, 1.5, ..., 7.5, 8 \quad \text{FWHM}$} and planet brightnesses e.g. \mbox{$f_p/f_* = 5, 5.5, ..., 10.5, 11 \quad \text{mag}$} into the raw dataset. We use the unsaturated PSF scaled by $f_p/f_*$ and potential attenuation due to a coronagraph as the planetary signal. 
\begin{figure}[t!]
	\epsscale{1.23}
	\plotone{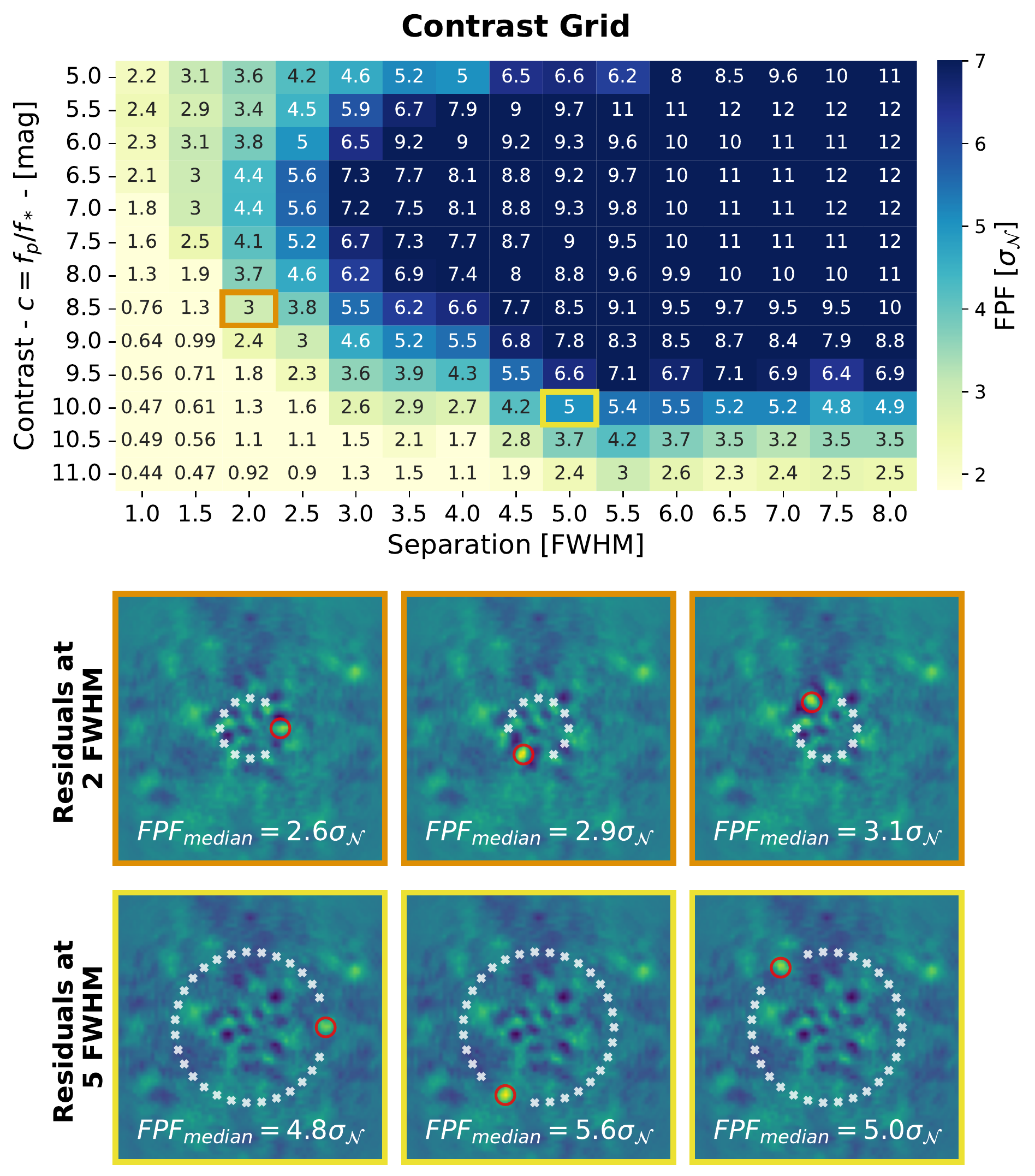}
	\caption{Detection limits for the $\beta$-Pictoris dataset under the assumption of Gaussian noise. The results under the assumption of Laplacian noise are presented later. The contrast grid (top pannel) gives the detection uncertainty (FPF) as a function of separation and contrast. Each value in the grid is based on 6 fake planet residuals. Below three example residuals are shown which were used to compute the contrast grid at the positions marked in orange and yellow. The $\text{FPF}_{\text{median}}$ given for each residual is the median FPF over different noise sample positions (see details in \mbox{Section \ref{sec:how_to_detection}}). The mean of all six $\text{FPF}_{\text{median}}$ values gives the final FPF value of the contrast grid. An interactive version of the plot is available on the documentation page of \texttt{applefy}: \url{https://applefy.readthedocs.io/}.}
	\label{fig:contrast_grid}
\end{figure}
In order to account for azimuthal variations we insert 6 artificial planets for each separation, one every 60 degrees. Some post-processing algorithms like the subtraction-based half-sibling regression presented in \cite{gebhardHalfsiblingRegressionMeets2022} make use of spatial correlations in the data. If we insert multiple planets at the same time these methods might learn to subtract the planet based on the movement of other planets in the data. Therefore, we insert only one planet at a time. For every inserted fake planet we run the data post-processing, in our case PCA, and save the residual images. Next, we estimate the FPF for each fake planet using the procedure explained in \mbox{Section \ref{sec:how_to_detection}}. Depending on the type of noise we can choose between the t-test or parametric bootstrapping.
The contrast grid, shown in \mbox{Figure \ref{fig:contrast_grid}}, combines all FPF estimates into single two dimensional grid.
The $5 \sigma_{\mathcal{N}}$ and $3 \sigma_{\mathcal{N}}$ contrast curves can be obtained by thresholding and interpolation of the grid.
As shown in the figure, no fake planet within 2 FWHM exceeds the $5 \sigma_{\mathcal{N}}$ detection threshold, no matter how bright it is. The reason for this is planet self-subtraction: At close separations a large fraction of the planetary signal is subtracted by the PCA noise model. If the planet gets brighter a progressively larger fraction of the planet flux is subtracted. In extreme cases an increase in planet brightness does not lead to a stronger signal in the residual. A detailed discussion of the effect is given in \mbox{Appendix \ref{sec:analytical_contast_curves}}.

The number of PCA components used during data post-processing affects the achieved contrast. While a higher number of PCA components leads to a stronger reduction of the noise, it also causes more loss of planetary signal. For the given dataset fewer components give higher contrast at close separations and more components provide better results for large separations. We recommend to compute contrast curves for a range of PCA components and report the overall best \citep{xuanCharacterizingPerformanceNIRC22018}. Our \texttt{python} package \texttt{applefy} provides the code needed to automate and parallelize these computations.

Under additional assumptions and for some data post-processing algorithms it is possible to compute the contrast curve analytically. This way the computation time can be reduced significantly. Current implementations in HCI packages like VIP \citep{gonzalezVIPVortexImage2017} or PynPoint \citep{stolkerPynPointModularPipeline2019} make use of the alternative. Also \texttt{applefy} allows to compute analytical contrast curves. A detailed discussion of this approach and how it can be used with parametric bootstrapping is presented in \mbox{Appendix \ref{sec:analytical_contast_curves}}.

\section{Results and Discussion}
\label{sec:results_and_discussion}

The parametric bootstrap presented in \mbox{Section \ref{sec:beyond_gaussian_noise}} quantifies planet detections for different types of residual noise. This way we can calculate the FPF of a potential planet candidate or, conversely, fix the FPF in order to determine detection limits. But, how accurate is the calculated FPF for non-Gaussian noise? Can we afford to make the wrong assumption? 

% Setup of the MC experiment
To answer these questions we run four Monte Carlo simulations considering Gaussian noise with ($\mu = 5$, $\sigma = 5$), Laplacian noise with ($\mu = 5$, $b = 5$) at 2 FWHM and 8 FWHM separation from the star. The separation in this simulation only influences the number of available noise values. Each Monte Carlo simulation is based on $10^9$ experiments for which we sample $n$ values from the respective noise distribution representing the planet signal under $H_0$ and the $n-1$ noise values available at 2 FWHM ($n = 11$) and 8 FWHM ($n=49$) separation. We run three tests to calculate the FPF for each experiment of the Monte Carlo simulation: The t-test, a parametric bootstrap under the assumption of Gaussian noise and a parametric bootstrap under the assumption of Laplacian noise. For the parametric bootstrap we use precomputed lookups as discussed in \mbox{Section \ref{sec:pivot}} with $B=10^8$. 

\begin{figure}[t!]
	\epsscale{1.2}
	\plotone{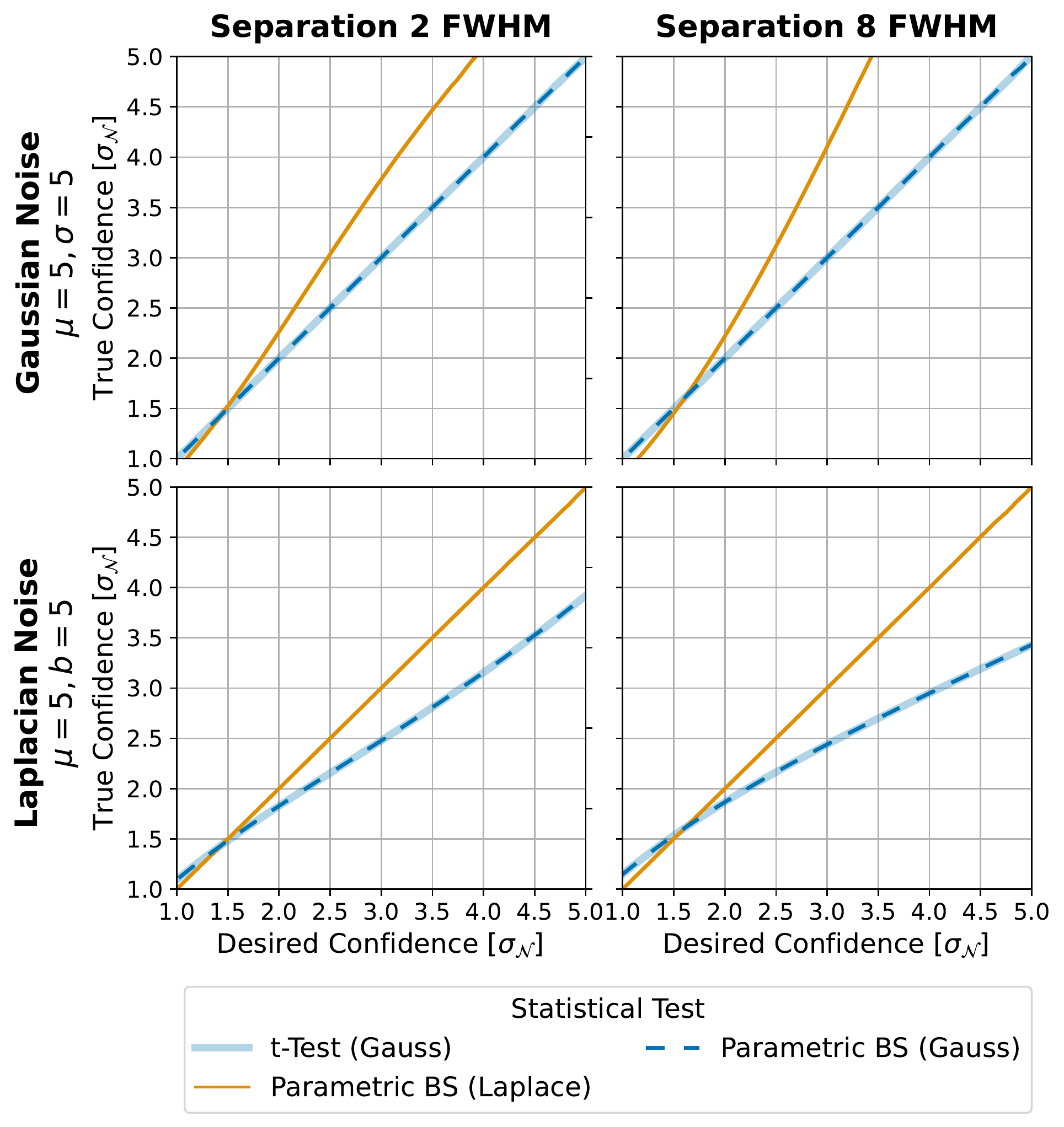}
	\caption{Effect of different noise distributions on the detection uncertainty. The plots give the relationship between the selected detection threshold and the actually observed FPF for the t-test and the parametric bootstrap tests presented in this paper. Any deviation from the diagonal corresponds to an over- or underestimation of the FPF. The results are based on a Monte Carlo simulation described in the text.}
	\label{fig:mc_sim_para_bs}
\end{figure}

We set different detection thresholds and count the number of experiments for which the tests are \emph{fooled} by the noise, that is the calculated FPF is smaller than the chosen threshold. This way we obtain the true FPF of the tests for a given detection threshold. A comparison of the true FPF with the FPF promised by the tests is shown in  \mbox{Figure \ref{fig:mc_sim_para_bs}}.
% Figure of convergence distribution for both Laplacian and Gaussian noise
We observe that if the simulated noise is Laplacian the obtained confidence for the t-test is substantially lower than the desired confidence. For example at 2 FWHM, a chosen confidence of $5 \sigma_{\mathcal{N}} = 2.867 \times 10^{-7} \quad \text{FPF}$ results in $3.8 \sigma_{\mathcal{N}} = 7.235 \times 10^{-5} \quad \text{FPF}$. This means that we underestimate the number of false positives by a factor of 250. Unlike the t-test, the Laplacian parametric bootstrap computes the correct FPF. If we suspect that our residual noise distribution is better described by a Laplacian we should choose the parametric bootstrap over the t-test. 
If the noise is Gaussian the t-test is well calibrated and provides accurate FPF results. The Laplacian parametric bootstrap, however, overestimates the FPF and the results become too pessimistic. This means that accurate knowledge about the underlying noise distribution is critical to determine the FPFs. The parametric bootstrap under the assumption of Gaussian noise is consistent with the t-test. This result demonstrates that bootstrapping is able to account for non-Gaussian noise and at the same time for the small sample statistics. 

\paragraph{Required Signal-to-Noise Ratio}
% How much SNR is needed for a 5 sigma detection Plot. 
\begin{figure}[t!]
	\epsscale{1.1}
	\plotone{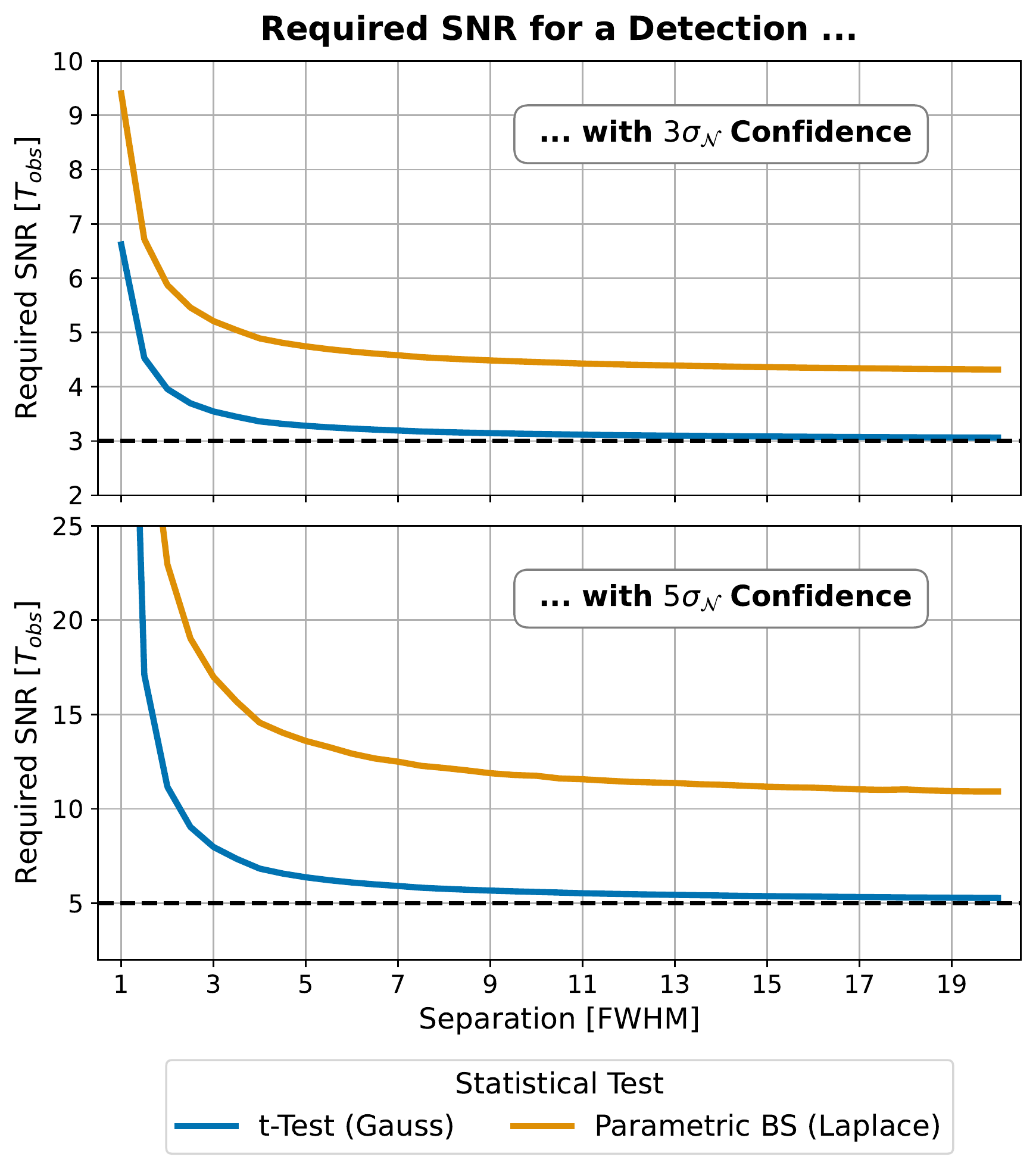}
	\caption{Relationship between the value of the test statistic $T_{\text{obs}}$ and the detection uncertainty specified as \mbox{$5 \sigma_{\mathcal{N}} = 2.87 \times 10^{-7} \quad \text{FPF}$} and \mbox{$3 \sigma_{\mathcal{N}} = 1.35 \times 10^{-3} \quad \text{FPF}$}. The t-distribution shown in blue converges towards a Gaussian for large separations (black dashed line) while the Laplacian parametric bootstrap remains heavier-tailed.}
	\label{fig:requried_SNR}
\end{figure}
For many applications, such as the computation of contrast curves, we are not interested in the FPF of a potential planet. Instead, we want to fix the FPF to constrain which value of the test statistic $T_{\text{obs}}$ (signal-to-noise ratio) is needed to achieve the desired confidence (FPF). For Gaussian noise this can be done by solving \mbox{Equation \ref{eq:FPF_ttest}} for $T_{\text{obs}}$. For Laplacian noise we use the procedure explained in \mbox{Appendix \ref{sec:analytical_contast_curves}}. The results for different separations are summarized in \mbox{Figure \ref{fig:requried_SNR}}.
%
% Final discussion of the result
As shown in the plot a significantly larger value of $T_{\text{obs}}$ is required under Laplacian noise. This is because the Laplacian distribution has heavier-tails and the occurrence rate of large noise values is higher.
If we aim for a detection confidence of $5 \sigma_{\mathcal{N}}$ the signal of the planet needs to be more than two times brighter compared to the limits of Gaussian noise. The effect is important irrespective of the separation from the star. This means that even if we have a large sample size, we are not robust to non-Gaussian noise. This is due to the noise at the position of the planet and the fact that the sample of the signal contains only one observation (see discussion in \mbox{Section \ref{sec:what_is_a_detection}}). For a detection threshold of $3 \sigma_{\mathcal{N}}$ the difference between Gaussian and Laplacian noise is less important but still not negligible. At large separations the t-test converges to the classical $5 \sigma$ or $3 \sigma$ limits. Both tests account for the effect of small sample statistics at separations close to the star.

\begin{figure}[t]
	\epsscale{1.1}
	\plotone{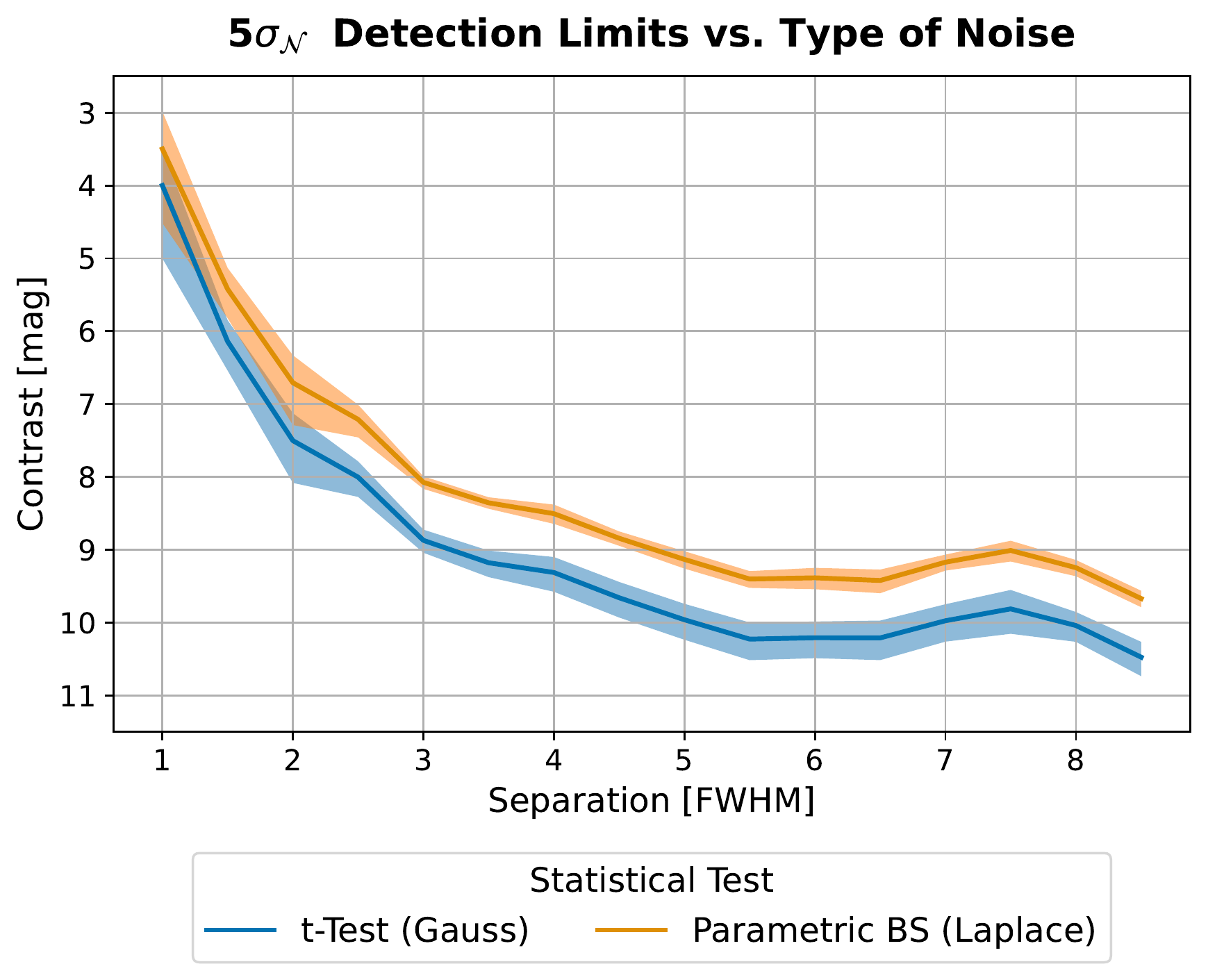}
	\caption{Comparison of the detection limits for the $\beta$-Pictoris datasets, once under the assumption of Gaussian (blue line) and once for Laplacian (orange line) residual noise. The solid line is the median contrast over 360 different placements of the noise values in the residual. The shaded area gives the mean absolute deviation from the median. We used 30 PCA components for all separations.}
	\label{fig:contrast_curves}
\end{figure}

\paragraph{Detection Limits}
% Goal: Compare the effect of non-Gaussian noise on detection limits
In order to investigate the effect of non-Gaussian noise on the detection limits, we compute contrast curves under the assumption of Gaussian and Laplacian noise. 
%
% Approach: Compute the exemplary for the Beta Pic dataset
For this purpose, we compare the results of the classical t-test with those of the Laplacian parametric bootstrap test. We follow the procedure discussed in \mbox{Appendix \ref{sec:analytical_contast_curves}} to calculate our contrast curves. We overcome the limitations discussed in \mbox{Section \ref{sec:critical}} i.e. we use spaced pixel instead of apertures to guarantee independent noise observations and take the median contrast over several different noise positions.
%
% Results are shown in the Figure
The results for the $\beta$-Pictoris dataset are shown in \mbox{Figure \ref{fig:contrast_curves}}.
%
% The key results: We are off by one magnitude
For separations $>2$ FWHM the contrast of the t-test is about one magnitude deeper compared to the contrast of the parametric bootstrap. This difference corresponds to a factor 2.5 in planetary brightness. At separations close to the star (1-2 FWHM) the difference between the two limits is smaller. At these separations the effect of small sample statistics becomes relevant. The t-test accounts for this effect through the heavier tails of the t-distribution. These heavier tails also partially mitigate the problem caused by non-Gaussian noise.
%
% Implications for the field in general
If the true noise is heavy-tailed, the contrast curve of the t-test is too optimistic. In such situations direct imaging surveys might have ruled out regions of the parameter space where we might still find planets. Note, non-Gaussian noise has a systematic effect on the results. That is, the error does not average out if several datasets are combined within a survey. If the datasets within one survey are all affected by heavy tailed speckle noise the whole survey is biased. 
%
% We do not know the true distribution of the noise
In reality, the true distribution of the noise is influenced by many factors and therefore often unknown. Hence, we cannot decide which of the two detection limits is actually correct.
Based on the related work discussed in Section \ref{sec:non_gaussian_noise} we would expect the noise in the speckle dominated regime, i.e. close to the star, to be better described by a Laplacian. At these separations, the true contrast is likely closer to the contrast curve of the parametric bootstrap under the assumption of Laplacian noise. At larger separations, the noise becomes more Gaussian and the results of the t-test or Gaussian parametric bootstrap are likely accurate.
It is important to note that we can not prove if the noise is sufficiently normal. A discussion about this problem is given in \mbox{Appendix \ref{sec:testing_gaussian}}.
Thus, as long as no additional knowledge about the noise is available, we have to accept that our contrast curves are potentially inaccurate by about one magnitude.

\section{Summary and Conclusions}
\label{sec:conclusion}
In this paper we presented a new framework to quantify exoplanet detections in high-contrast imaging. Through the use of parametric bootstrapping, our new metric is able to estimate robust detection limits for any type of residual noise. Our metric, however, assumes that the noise distribution is known. A comparison of the detection limits under the assumption of Gaussian and Laplacian noise revealed that commonly used metrics, such as the t-test, might be too optimistic in case of speckle dominated observations. 
The occurrence rate of large noise values is higher in case of Laplacian noise compared to Gaussian noise. This results in a higher risk to obtain a high signal-to-noise value originating from the noise. For example, the risk that the noise produces a false detection at 2 FWHM distance from the star with a signal-to-noise ratio of 5 is about 250 times higher under Laplacian than under Gaussian noise. Therefore, the signal-to-noise ratio should not be considered as a direct measure for the detection uncertainty. Only if we take the sample size and the correct noise distribution into account the signal-to-noise ratio becomes interpretable. 

The link between the detection limit and the noise characteristics makes a fair comparison between HCI observations difficult. This is especially the case if the noise distributions differ between the datasets we want to compare. An example for this is the development of new post-processing algorithms. If we compare algorithms that produce residuals with different noise, the comparison of the methods is likely biased. The same applies for any comparison between observations taken under different circumstances: e.g. ground-based vs. space-based observations, different instruments or observing strategies such as ADI and RDI. If we want to compare inhomogeneous data, we have to take possible biases arising from statistics into account.
We recommend to compute detection limits under different assumptions to set optimistic and conservative bounds for the achieved contrast. Studies on the residual noise using for example Q-Q plots can provide valuable insights about the noise. But they can never prove that the noise is sufficiently normal to use a t-test. Apart from graphical tools, quantitative tests such as the Shaphiro-Wilk test can only reject normality. They can never proof that the data is actually normal.

Future work should seek to better understand the speckle statistics of the HCI residuals. If we knew the true distribution of noise as a function of observing conditions, instrument, and data post-processing, we could use the bootstrapping algorithm presented in this paper to determine the true contrast. Alternatively, we can use non-parametric methods such as the non-parametric bootstrap to estimate the statistics directly from the data. Future work should further investigate the origin of non-Gaussian noise and develop new methods to account for the spatial dependencies of speckle noise.

%\newline \newline\newline

\acknowledgments{We thank the anonymous referee for a critical and constructive review of the original manuscript which helped improve the quality of the paper significantly. 

This work was supported by an ETH Zurich Research Grant. M.J.B. and S.P.Q. gratefully acknowledge the financial support from ETH Zurich. Parts of this work has been carried out within the framework of the National Centre of Competence in Research PlanetS supported by the Swiss National Science Foundation (SNSF) under grants 51NF40\_182901 and 51NF40\_205606. Parts of this work were supported by the SNSF via grant number 200020\_200300. S.P.Q., E.O.G., F.A.D., and J.H. acknowledge the financial support of the SNSF. O.A. acknowledges funding from F.R.S.-FNRS, and from the European Research Council (ERC) under the European Union's Horizon 2020 research and innovation programme (grant agreement No 819155). G.C. thanks the Swiss National Science Foundation for financial support under grant number P500PT\_206785.

\textit{Author contributions}: M.J.B. carried out the main analyses and wrote the manuscript. He further programmed the publicly available the python package \texttt{applefy} and wrote the documentation page on ReadTheDocs. E.O.G. proposed the initial idea to explore parametric bootstrapping and pivoting to address the detection problem. T.D.G. contributed to the programming of \texttt{applefy}. He and G.C. helped to debug the alpha version of the code. All authors discussed the results and commented on the manuscript.}

\bibliography{Metrics_Part_1_literature}{}
\bibliographystyle{aasjournal}

\appendix
\newpage
\section{Aperture placement}
\label{sec:rotation}
% Aspect 1: Implementation Details --------------------------------------------------------------

\begin{figure}[t!]
	\epsscale{1.15}
	\plotone{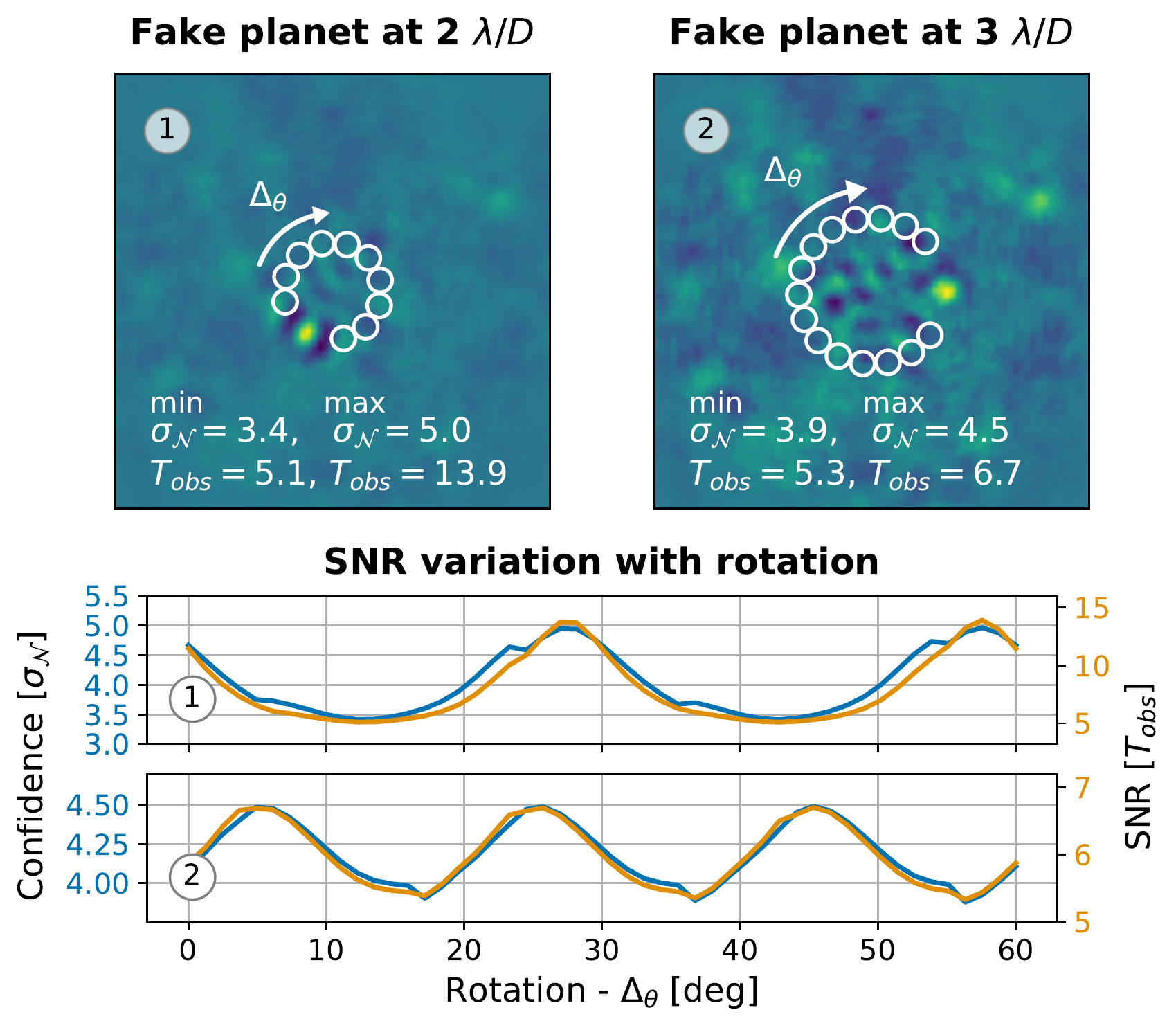}
	\caption{The effect of the aperture placement on the detection uncertainty. The top two images show the residuals with the inserted fake planets at 2 and 3 $\text{FWHM}$. The white circles indicate the initial positions of the apertures used to extract the noise. Below we show the computed test statistic $T_{\text{obs}}$ (orange) and detection uncertainty / FPF (blue) for different rotations of the initial noise positions. The difference between the values of $T_{\text{obs}}$ and the FPF are due to the number of apertures changing by $\pm 1$ during the rotation.}
	\label{fig:rotation}
\end{figure}

The first step of the hypothesis test illustrated in \mbox{Figure \ref{fig:hypothesis_testing}} is the extraction of the signal and the noise. The position of the signal can be estimated by maximizing the flux inside the signal aperture. The positions for the noise are commonly selected depending on the final signal position. This choice, however, is arbitrary and any other arrangement of non-overlapping apertures should yield similar results.
In order to check if $T_{\text{obs}}$ is invariant to the positioning of the apertures we insert fake planets into the \mbox{$\beta$ Pictoris} dataset and perform a PCA-based PSF subtraction. We insert one fake planet at a separation of \mbox{$2 \text{ FWHM}$} and one at \mbox{$3 \text{ FWHM}$}. 
We measure the flux of the inserted fake planets $\overline{Y_1}$ in each residual image and extract values for the noise $\overline{X_1}, ..., \overline{X_n}$ by using apertures with the same separation from the star. At \mbox{$2 \text{ FWHM}$} we have $n = 12$ and at \mbox{$3 \text{ FWHM}$} we have $n = 18$ apertures available. Apertures close to the planet signal are excluded in order to account for self-subtraction artefacts next to the planet. This reduces $n$ by about 2 apertures. Next, we compute $T_{\text{obs}}$ and the FPF as explained in \mbox{Section \ref{sec:what_is_a_detection}}. We repeat the calculation several times for slightly rotated aperture positions. In total we rotate all apertures by 60 degrees which corresponds to a displacement by $60/360 \cdot n = 2$ apertures at \mbox{$2 \text{ FWHM}$} and 3 apertures at \mbox{$3 \text{ FWHM}$}. The results are summarized in \mbox{Figure \ref{fig:rotation}}.

As shown in the plot the computed FPF changes significantly depending on the rotation of the aperture positions $\Delta_{\theta}$. Especially at $2 \text{ FWHM}$ the values of $T_{\text{obs}}$ vary by a factor of 2. The corresponding confidence levels span from $3.4 \sigma_{\mathcal{N}}$ to $5.0 \sigma_{\mathcal{N}}$. At \mbox{$3 \text{ FWHM}$} the effect is smaller but still not negligible.
The position of the apertures influences whether speckles end up between or inside the apertures. If a bright speckle is located between two apertures, its flux will be split and $\hat{\sigma}_{\mathcal{X}}$ decreases. Vice versa, if the speckle is in the center of an aperture, $\hat{\sigma}_{\mathcal{X}}$ will be large. At small separations only a few apertures are available and a single speckle can already have a large effect on $T_{\text{obs}}$.

We recommend to estimate $T_{\text{obs}}$ several times for slightly rotated aperture positions. Instead of 60 degrees total rotation a rotation by $360 / (2 \pi r)$ degrees is sufficient for a displacement by one aperture (one oscillation; compare \mbox{Figure \ref{fig:rotation}}). The results can be summarized by reporting the median value of $T_{\text{obs}}$ over all rotations and the median absolute deviation (MAD) as a measure for the influence of the aperture placement. The minimum and maximum value of $T_{\text{obs}}$ give the worst- and best-case result. The same procedure should be used if spaced pixel are used instead of apertures (see discussion in \mbox{Section \ref{sec:independence}}.
\section{Analytical Contrast Curves}
\label{sec:analytical_contast_curves}
\begin{figure}[b!]
	\epsscale{1.1}
	\plotone{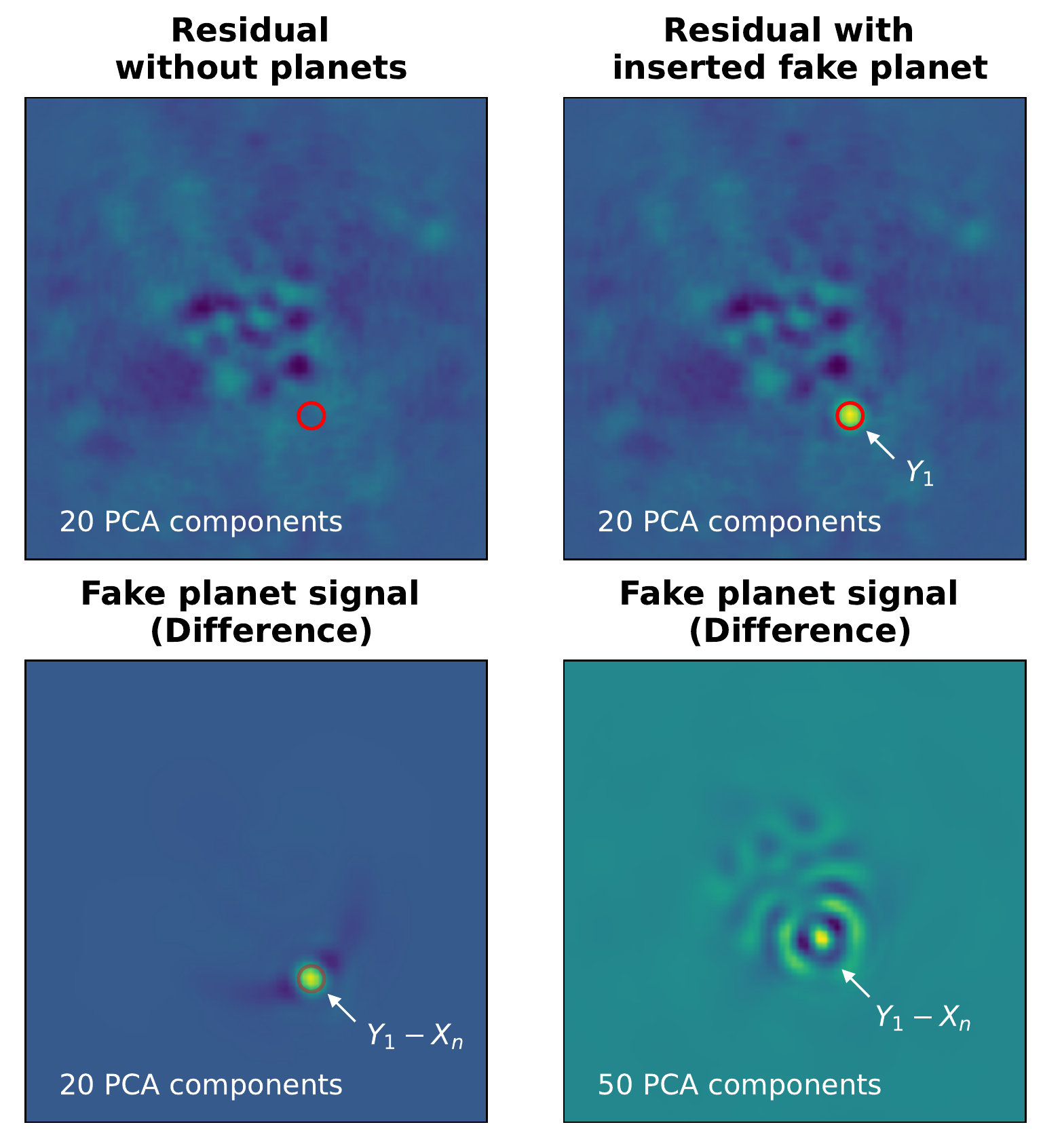}
	\caption{The top two panels show example residuals for the $\beta$-Pictoris dataset with and without fake planets. The bottom left image gives their difference. The bottom right image shows the result of the same experiment but for a fake planet closer to the star and with 50 PCA components. For 20 PCA components the presence of the planet only affects values of the residual in the direct neighborhood of the signal. For 50 PCA components the whole residual image changes. This is potentially problematic as discussed in the text.}
	\label{fig:fake_planets}
\end{figure}
\begin{figure}[t!]
	\epsscale{1.2}
	\plotone{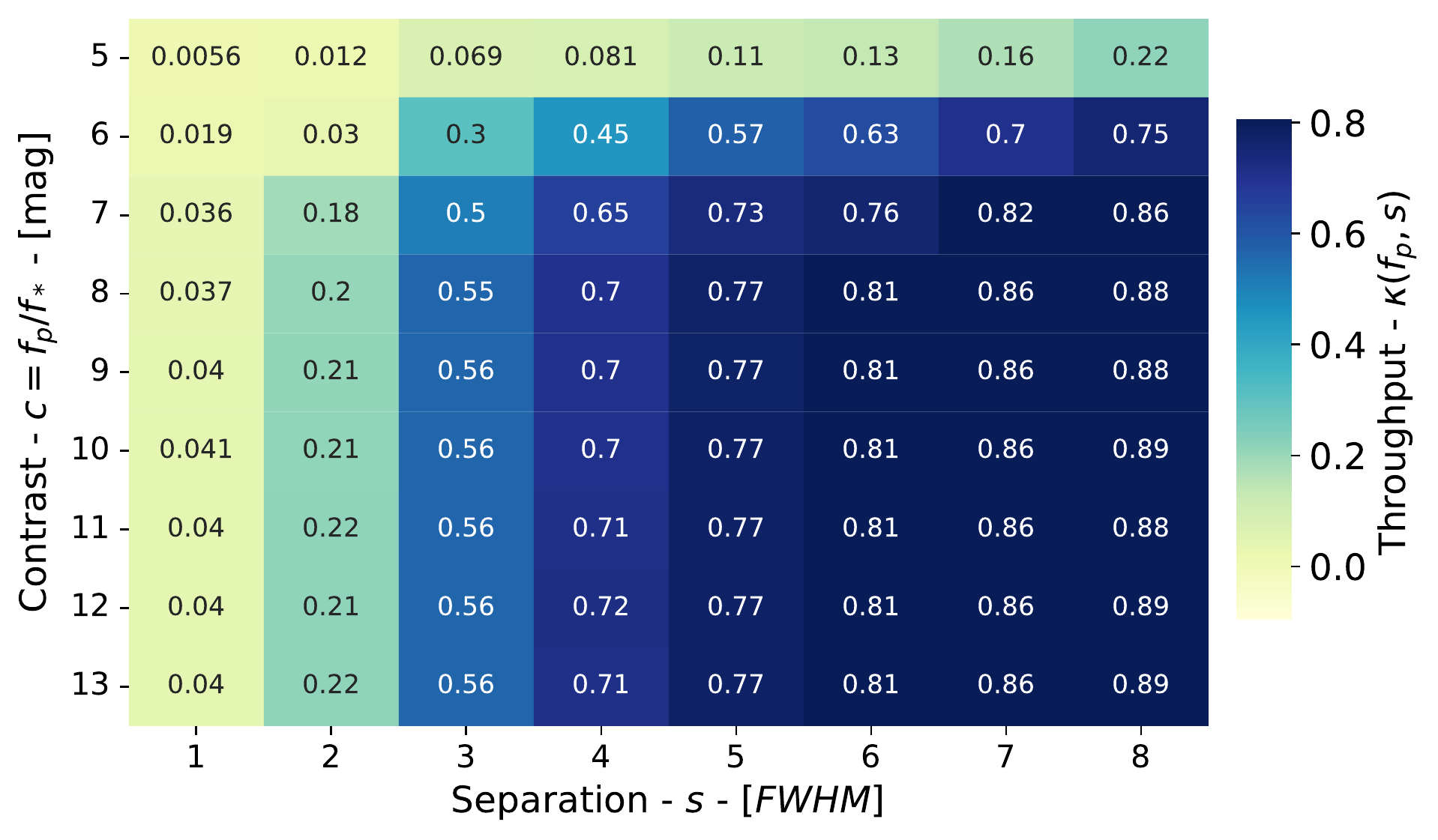}
	\caption{Throughput of the $\beta$ Pictoris dataset for 30 PCA components. Every value in the plot is the average of six experiments with artificial planets inserted at 6 different azimuthal positions. }
	\label{fig:throughputs_beta_pic}
\end{figure}

\begin{figure}[t!]
	\epsscale{1.1}
	\plotone{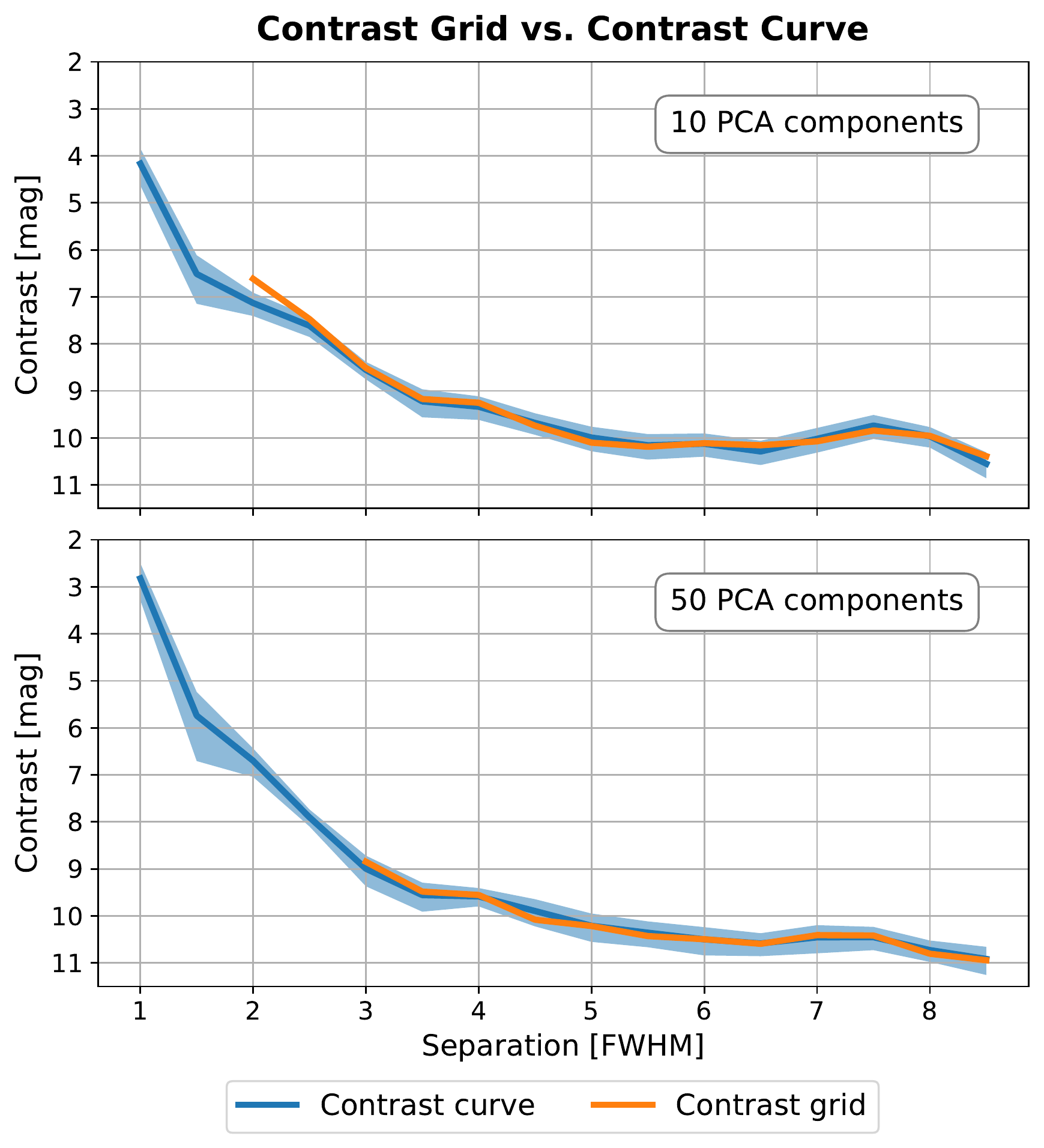}
	\caption{Comparison of the analytical contrast curve (based on the approximation that the throughput is only a function of separation) with the complete contrast grid presented in the main paper \mbox{(Section \ref{sec:detection_limits})}. The orange line gives the contrast grid thresholded at $5 \sigma_{\mathcal{N}}$. The results are based on the assumption of Gaussian noise.}
	\label{fig:contrast_curve_vs_map}
\end{figure}

The calculation of one contrast grid presented in \mbox{Section \ref{sec:detection_limits}} requires to process several hundred datasets with inserted fake planets. This computation can be very time consuming. Under some mild assumptions, it is possible to reduce this computation time considerably. This is the commonly used approach of how contrast curves are calculated in packages like \texttt{PynPoint} \cite{stolkerPynPointModularPipeline2019} or \texttt{VIP} \cite{gonzalezVIPVortexImage2017}. We start by rearranging \mbox{Equation \ref{eq:SNR}}

\begin{equation}
	\label{eq:needed_flux_residual}
	Y_1 = T_{\text{obs}} \cdot \hat{\sigma}_{\mathcal{X}} \sqrt{1 + \frac{1}{n}} + \hat{\mu}_\mathcal{X} \, .
\end{equation}
The value of $Y_1$ is the brightness of the signal required in the residual to reach $T_{\text{obs}}$. Given a detection threshold specified as a FPF we can derive the $T_{\text{obs}}$ needed to be counted as a detection. For Gaussian noise this can be done by solving \mbox{Equation \ref{eq:FPF_ttest}} for $T_{\text{obs}}$ were $p(T = t| H_0)$ is given by the t-distribution. For non-Gaussian noise, larger values of $T$ might be required to reach the same FPF (see \mbox{Figure \ref{fig:requried_SNR}}). We re-use the sorted bootstrap results \mbox{$T_{(1)}^* \leq ... \leq T_{(B)}^*$}, discussed in step 6 of \mbox{Section \ref{sec:para_bs}}, and estimate $T_{\text{obs}}$ by linear interpolation:
	 
\begin{equation}
\label{eq:constrain_T}
	 	T_{\text{obs}} = T^*_{(\lfloor{a}\rfloor)} (a - \lfloor{a}\rfloor) + T^*_{(\lceil{a}\rceil)} (\lceil{a}\rceil - a)
\end{equation}
where $a = (B - 1) (1 - \text{FPF})$ gives the index of the two bootstrap results $T^*_{(\lfloor{a}\rfloor)}$ and $T^*_{(\lceil{a}\rceil)}$, which are closest to the required FPF. This step is the inverse of the linear interpolation explained in step 7 of \mbox{Section \ref{sec:para_bs}}. Due to planet over- and self-subtraction during the data post-processing, the flux of the planet in the residual is attenuated. We can describe this effect by

\begin{equation}
	\label{eq:residual_signal}
	Y_1 = f_p \cdot \kappa(f_p, s) + X_{n+1}
\end{equation}
were $X_{n+1}$ is the speckle noise at the position of the planet and $\kappa(f_p, s) \in [0, 1]$ is the throughput accounting for the attenuation of the data post-processing and potential coronagraphs. The separation is denoted as $s$. The throughput can be computed using the following procedure:
\begin{itemize}
	\item As for the contrast grid in \mbox{Section \ref{sec:detection_limits}}, insert artificial planets at different separations $s$ with different contrast $c = f_p / f_*$ into the raw data. Run the data post-processing to compute their residuals. In order to account for azimuthal variations we insert six planets, one at a time, for each separation and contrast.
	\item Compute one residual without fake planets. 
	\item Subtract the planet-free residual from every fake planet residual and estimate the flux at the position of the fake planet. Thanks to the linearity of PCA this gives us $Y_1 - X_{n+1}$. Since our statistic is based on pixel spaced by one FWHM, we integrate the flux within an area of one pixel around the position of the planet (compare \mbox{Section \ref{sec:how_to_detection}}). We note, that this step is only valid for sufficiently faint planets which do not affect the PCA component matrix. It further does not hold for all existing post-processing techniques.
	\item Use \mbox{Equation \ref{eq:residual_signal}} to compute the throughput.
\end{itemize}
Examples for residuals with and without fake planets are shown in \mbox{Figure \ref{fig:fake_planets}}. The throughput is summarized in \mbox{Figure \ref{fig:throughputs_beta_pic}}. It depends on the brightness of the inserted fake planet $f_p$ as well as its separations from the star. For faint planets, the PCA basis is not changed by the presence of the planet and the throughput converges. Bright planets, on the other hand, influence the PCA basis and cause additional signal loss. Planets which are close to the detection limit are usually faint. That is, under the assumption that these planets do not affect the PCA basis we can simplify $\kappa(f_p, s) = \kappa(s)$ and use the convergence throughput. That is, we only need to compute the last row of \mbox{Figure \ref{fig:throughputs_beta_pic}}. The limit for the planet to star contrast can be calculated by:

\begin{equation}
\label{eq:contrast}
	c = \frac{T_{\text{obs}} \cdot \hat{\sigma}_{\mathcal{X}} \sqrt{1 + 1/n} + \hat{\mu}_\mathcal{X} - X_{n+1}}{\kappa(s) \cdot f_*}
\end{equation}
The values of $X_{n+1}$, $\hat{\sigma}_{\mathcal{X}}$ and $\hat{\mu}_\mathcal{X}$ are based on noise observations from the planet free residual. They are again dependent on where the noise gets extracted. We propose the following procedure to account for this effect:
\begin{enumerate}
	\item Estimate $f_*$ on the unsaturated PSF by integration of the flux within an area of one pixel around the star.
	\item Choose a detection threshold, for example $5 \sigma_{\mathcal{N}} = 2.87 \times 10^{-7} \text{FPF}$.
	\item For each separation extract noise values  spaced by 1 FWHM from the planet free residual. Use one value as $X_{n+1}$ and the rest as $\mathcal{X}$.
	\item Make an assumption about the noise to constrain $T_{\text{obs}}$. For Gaussian noise solve \mbox{Equation \ref{eq:FPF_ttest}} for $T_{\text{obs}}$. For non-Gaussian noise use \mbox{Equation \ref{eq:constrain_T}}.
	\item Use \mbox{Equation \ref{eq:contrast}} to compute the contrast.
	\item Repeat the steps 3.-5. with different noise positions (compare \mbox{Figure \ref{fig:rotation}}). Report the median contrast over all repetitions. The mean absolute deviation from the median can be used as a measure for the uncertainty introduced by the placement of the noise positions.
\end{enumerate}
The given procedure assumes that the planet signal has no effect on the noise sample $\mathcal{X}$. As shown in the bottom right plot in \mbox{Figure \ref{fig:fake_planets}} this is not necessary the case for high number of PCA components. Under such circumstances, the calculation of a complete contrast grid is favorable. A comparison of the analytical contrast curve with the results of the contrast grid is shown in \mbox{Figure \ref{fig:contrast_curve_vs_map}}.
As shown in the figure, both methods are consistent. In other words, the detection uncertainty of the artificial planets inserted for the contrast grid agrees with the contrast curve. At small separations, the contrast curve reaches a regime where the PCA basis is changed by the planet. For these separations no $5 \sigma_{\mathcal{N}}$ contrast exist, meaning that no planet, no matter how bright, will ever give a $5 \sigma_{\mathcal{N}}$ detection. This effect can only be identified with the contrast grid.

\section{Testing for non-Gaussian noise}
\label{sec:testing_gaussian}

The noise distribution of HCI residuals can vary within and between observations. To confirm that the noise is sufficiently normal for the t-test to be applicable, various tests are carried out prior to the calculation of the detection limits. For example \cite{mawetFUNDAMENTALLIMITATIONSHIGH2014} and \cite{ottenONSKYPERFORMANCEANALYSIS2017} compute histograms based on pixel values within annuli of $1 \lambda / D$ width. A visual inspection of histograms can provide first insights about the general distribution shape. A direct comparison with the normal distribution, however, is difficult. This is especially the case for the tails of the distribution, i.e. the occurrence rate of bright events. These events are of special importance for the calculation of the detection limits. Q-Q plots are a good alternative to histograms as they allow to directly compare the noise with the normal distribution. In this way, insights about a possible over- or underestimation of the contrast can be obtained.

\begin{figure}[t!]
	\epsscale{1.15}
	\plotone{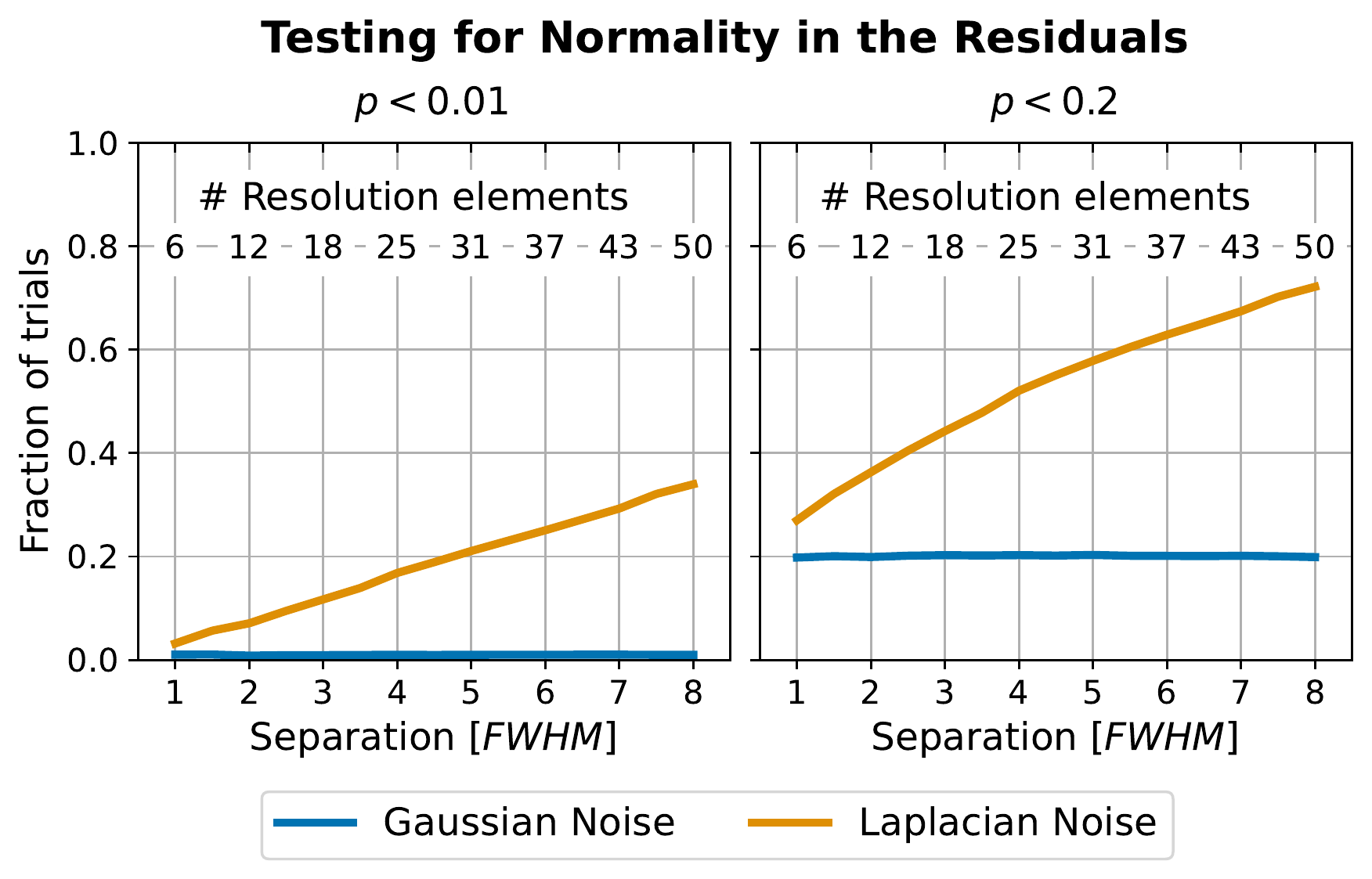}
	\caption{Sensitivity and false positive rate (FPR) of the Shapiro-Wilk test as a function of separation from the star. For each separation, $10^5$ samples are drawn, each containing as many values as independent resolution elements are available. We sample from a Gaussian distribution to measure the FPR (blue) and from a Laplacian distribution to measure the sensitivity (orange). The left and right plots show the results for two different thresholds $p < 0.01$ and $p < 0.2$.}
	\label{fig:shapiro}
\end{figure}

The Shapiro-Wilk test \citep{shaphiroAnalysisVarianceTest1965} can be used as a quantitative measure for normality. The test can be used to reject the null hypothesis that a sample was drawn from a normal distribution. For the test to be applicable, it is assumed that the values of the sample are independent. Due to the spatial size of the speckles, neighboring pixel in the HCI residuals are never independent. Thus, the test has to be calculated on pixel which are at least \mbox{1 FWHM} apart. Results computed on non-independent values \cite[see e.g.][]{absilSearchingCompanionsAU2013} are likely biased. The output of the test is a p-value corresponding to the remaining risk that the sample is compatible with the null hypothesis. For example, if we observe $p=0.01$, we know that $1\%$ of the observations under Gaussian noise have a value of the test statistic equal or more extreme than ours. By setting a threshold, we can specify the accepted error by which Gaussian noise is misclassified as non-Gaussian noise. Although the Shapiro-Wilk test has shown superior performance compared to other tests \citep{razaliPowerComparisonsShapirowilk2011}, it lacks power in detecting non-Gaussian noise at close separation to the star. We study this effect in \mbox{Figure \ref{fig:shapiro}}.

The highest true positive rate is reached for the threshold $p<0.2$ at 8 FWHM and is just about $60\%$. The results get worse for close separations. This is problematic as speckle noise is expected to be the dominant noise source close to the star. While sensitivity is one problem, preliminary testing comes with a "logical problem" \citep{scholzKSampleAndersonDarling1987}: "Because insufficient evidence exists to reject normality, normality will be considered true" \citep{rochonTestNotTest2012}. If we cannot reject the null hypothesis, we must not conclude that our data is normal, i.e., combined with the low sensitivity for small sample sizes, we can neither prove nor disprove normality. As concluded in \cite{rochonTestNotTest2012} "support for the assumption of normality must come from extra-data sources" \citep{easterlingEffectPreliminaryNormality1978}.

\section{Pivot for the Laplace Distribution}
\label{sec:pivot_proof}
Let $Z$ be a random variable following a standard distribution with no unknown parameters (e.g. a standard Gaussian or Laplacian). The random variables $X = Z q_\mathcal{X} + w_\mathcal{X} \sim \mathcal{F}(q_\mathcal{X}, w_\mathcal{X})$ and $Y = Z q_\mathcal{Y} + w_\mathcal{Y} \sim \mathcal{F}(q_\mathcal{Y}, w_\mathcal{Y})$ with ${w_\mathcal{X}, w_\mathcal{Y} \in \mathbb{R}}$ and ${q_\mathcal{X}, q_\mathcal{Y} \in \mathbb{R}^+}$ from a location-scale family of $Z$. In case of the normal distribution $Z \sim \mathcal{N}(0, 1)$ with $q = \sigma, t=\mu$ and $Z \sim \mathcal{L}(0, 1)$ with $q = b, w=\mu$ for the Laplacian distribution.

We consider a two sample dataset $\mathcal{X} = {X_1, ..., X_n} \sim \mathcal{F}(q_\mathcal{X}, w_\mathcal{X})$ and $\mathcal{Y} ={Y_1, ..., Y_m} \sim \mathcal{F}(q_\mathcal{Y}, w_\mathcal{Y})$ under the assumption of equal scale ${q_\mathcal{X} = q_\mathcal{Y} = q}$. We want to test the null hypothesis ${H_0: w_\mathcal{X} - w_\mathcal{Y} = 0}$ against the alternative ${H_1: w_\mathcal{Y} - w_\mathcal{X} > 0}$ using the test statistic of the two sample t-test (c.f. Equation \ref{eq:SNR} for the special case of $m=1$ used in HCI):

\begin{equation}
T = \frac{\hat{\mu}_\mathcal{Y} - \hat{\mu}_\mathcal{X}}{\hat{\sigma}_{\mathcal{X}, \mathcal{Y}} \sqrt{1/n + 1/m}}
\end{equation}
where $\hat{\mu}_\mathcal{X}$ and $\hat{\mu}_\mathcal{Y}$ are the sample averages:

\begin{equation}
	\hat{\mu}_\mathcal{X} = \overline{\mathcal{X}} =\frac{1}{n} \sum_{i=1}^{n} X_i \qquad \hat{\mu}_\mathcal{Y} = \overline{\mathcal{Y}} = \frac{1}{m} \sum_{j=1}^{m} Y_j
\end{equation}
and $\hat{\sigma}_{\mathcal{X}, \mathcal{Y}}$ is the pooled standard deviation of the two samples:

\begin{equation}
	\hat{\sigma}_{\mathcal{X}, \mathcal{Y}}^2 = \frac{(n-1)\hat{\sigma}_{\mathcal{X}}^2 + (m-1)\hat{\sigma}_{\mathcal{Y}}^2}{n+m-2}
\end{equation}
with

\begin{eqnarray}
	\hat{\sigma}_{\mathcal{X}}^2 &= \frac{1}{n-1}\sum_{i=1}^n \left(X_i - \overline{\mathcal{X}}\right)^2 \\ \hat{\sigma}_{\mathcal{Y}}^2 &= \frac{1}{m-1}\sum_{j=1}^m \left(Y_j - \overline{\mathcal{Y}}\right)^2
\end{eqnarray}
Under $H_0$ the test statistic $T$ follows a distribution that is independent of the parameters $w_\mathcal{X}, w_\mathcal{Y}, q$, i.e. it is a pivot. The following proof is inspired by exercise 9.9 in \cite{casellaStatisticalInference2002}.

\textbf{Proof:}
\begin{align*}
	\hat{\sigma}_{\mathcal{X}}^2 
	&= \frac{1}{n-1} \sum_{i=1}^n \left(Z_i q_\mathcal{X} + w_\mathcal{X} - \overline{Z} q_\mathcal{X} - w_\mathcal{X} \right)^2 \\
	&= \frac{q_\mathcal{X}^2}{n-1} \sum_{i=1}^n \left(Z_i - \overline{Z}\right)^2 \\
	&= q_\mathcal{X}^2 \hat{\sigma}_{Z_n}^2
\end{align*}
The same holds for $\hat{\sigma}_{\mathcal{Y}}^2 = q_\mathcal{Y}^2\hat{\sigma}_{Z_m}^2$. With $q_\mathcal{X} = q_\mathcal{Y} = q$ it follows:

\begin{align*}
	\hat{\sigma}_{\mathcal{X}, \mathcal{Y}}^2 
		&= \frac{q^2(n-1)\hat{\sigma}_{Z_n}^2 + q^2(m-1)\hat{\sigma}_{Z_m}^2}{n+m-2} \\
		&= q^2 \hat{\sigma}_{Z_n Z_m}^2
\end{align*}
We re-write the test statistic $T$:

\begin{align*}
	T &= \frac{\frac{1}{n} \sum_{i=1}^{n} X_i - \frac{1}{m} \sum_{j=1}^{m} Y_j}
			  {\hat{\sigma}_{\mathcal{X}, \mathcal{Y}} \sqrt{(1/n + 1/m)}}\\
		&= \frac{\frac{1}{n} \sum_{i=1}^{n} \left(Z_i q + w_\mathcal{X} \right)  - \frac{1}{m} \sum_{j=1}^{m} \left(Z_j q + w_\mathcal{Y}\right)}
		{\hat{\sigma}_{\mathcal{X}, \mathcal{Y}} \sqrt{(1/n + 1/m)}} \\
		&=\frac{q \left(\frac{1}{n}\sum_{i=1}^{n} Z_i - \frac{1}{m}\sum_{j=1}^{m} Z_j\right)}{q \hat{\sigma}_{Z_n Z_m} \sqrt{(1/n + 1/m)}} + \frac{w_\mathcal{X} - w_\mathcal{Y}}{q \hat{\sigma}_{Z_n Z_m} \sqrt{(1/n + 1/m)}}
\end{align*}
Under $H_0$ we have $w_\mathcal{X} = w_\mathcal{Y}$. It follows:

\begin{equation}
	T=\frac{\overline{Z_n} - \overline{Z_m}}{\hat{\sigma}_{Z_n Z_m} \sqrt{(1/n + 1/m)}}
\end{equation}
All  quantities $\overline{Z_n}, \overline{Z_m}, \hat{\sigma}_{Z_n Z_m}$ are independent of $w_\mathcal{X}, w_\mathcal{Y}, q$. $\blacksquare$
\section{Speckle Noise Correlation}
\label{sec:Correlation}

\begin{figure}[b!]
	\epsscale{1.13}
	\plotone{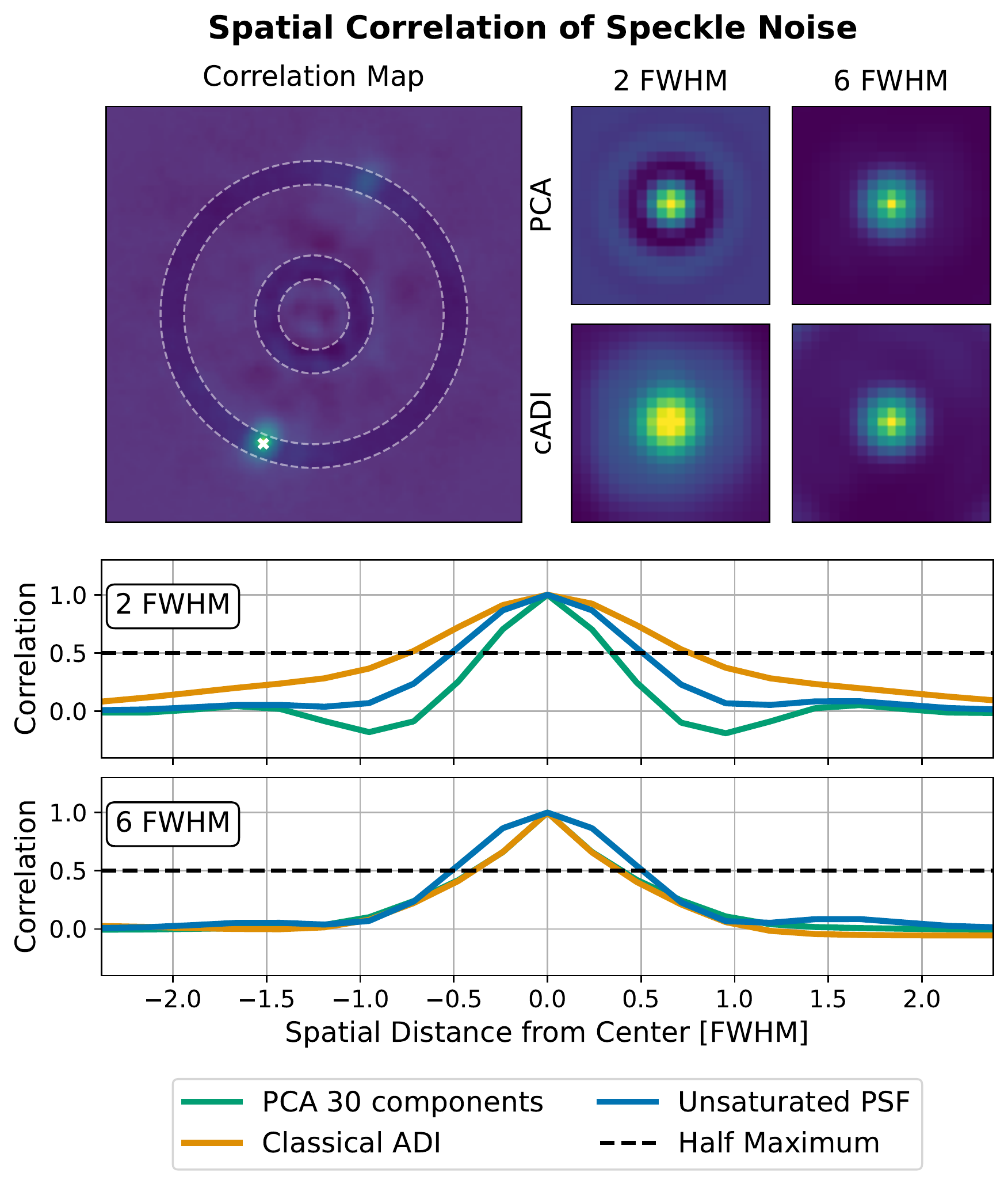}
	\caption{Estimates of the spatial noise correlations of the $\beta$-Pictoris dataset. The image on the top left is an example of a correlation map. It shows how much the pixel marked with the white cross is correlated w.r.t. all other pixel in the image. We compute one correlation map for every pixel in the residual. The top right images show the local averages of the correlation maps at \mbox{2 FWHM} and \mbox{6 FWHM} for cADI and PCA. The regions used to compute the averages are marked in the correlation map. The bottom two plots give the one-dimensional profile of the local correlation maps (top right plots) in comparison to the unsaturated PSF of the star.}
	\label{fig:noise_correlation}
\end{figure}

A key assumption of the hypothesis tests discussed in this paper is that the observations in $\mathcal{Y}$ and $\mathcal{X}$ are independent. This assumption is fundamental for both the t-test as well as the parametric bootstrap. In \mbox{Section \ref{sec:critical}}, we motivate that pixel values spaced by 1 FWHM are approximately independent. This, however, is only the case if the actual correlations in the data follow the shape of the unsaturated PSF. In the following, we compute the spatial correlations of the noise in the $\beta$ Pictoris dataset processed with PCA \citep{amaraPYNPOINTImageProcessing2012,soummerDETECTIONCHARACTERIZATIONEXOPLANETS2012} and cADI \citep{maroisAngularDifferentialImaging2006}:
\begin{enumerate}
	\item We start the analysis with the pre-processed stack of images in which the star is located in the center of the frames. For cADI we subtract the median frame from every image in the stack. For PCA we subtract the first 30 PCA components. We de-rotate the frames according to their parallactic angles. The result is the sequence of residual images, which we average to obtain the final residual image.
	\item We compute the correlation of every pixel along the temporal domain w.r.t. all other pixel. For this we use the sequence of de-rotated residuals after PSF-subtraction. This way we obtain a single two-dimensional correlation map for each pixel.
	\item We crop the correlation maps such that the pixel they correspond to is in the center of the cropped map. The resulting maps show the local correlations for each pixel in the residual frame. 
	\item We average the cropped correlation maps for all pixel within an annulus at \mbox{2 FWHM} and \mbox{6 FWHM} distance from the star.
\end{enumerate}
The results and a comparison with the unsaturated PSF of the star is shown in \mbox{Figure \ref{fig:noise_correlation}}.
We observe that the true correlations in the data do not always follow the shape of the PSF. Further, they are influenced by the separation from the star as well as the data post-processing used. For cADI the FWHM of the local correlation map at \mbox{2 FWHM} is about 60 \% larger compared to the PSF-FWHM. Similar results were previously observed by \cite{golombPlanetEvidencePlanetNoise2021} for data from the Gemini Planet Imager. If the spatial correlation length is large, a spacing of 1 PSF-FWHM might not be sufficient to ensure that the noise observations in $\mathcal{X}$ are independent. 
At \mbox{6 FWHM} the FWHM of the local correlations is slightly smaller than the PSF-FWHM. If we use PCA, the local correlations at \mbox{2 FWHM} change while they stay the same at \mbox{6 FWHM}. Future work should further investigate this behavior to better understand the spatial correlations of speckle noise in real data. A possible path for this could be bootstrapping for dependent observations \citep{zoubirBootstrapTechniquesSignal2004}.




\end{document}